\documentclass[11pt,times]{article}
\usepackage{amsmath,mathptmx,amssymb}
\usepackage{tikz}
\usepackage{graphicx} 
\usepackage {epstopdf}
\graphicspath{{Figures/}}
\usepackage{caption,subcaption}
\usepackage{enumitem}
\usepackage{pgfplots}
\usetikzlibrary{external}
\pgfplotsset{compat=newest}
\usetikzlibrary{plotmarks}
\usetikzlibrary{arrows.meta}
\usepgfplotslibrary{patchplots}
\usepackage{grffile}
\newtheorem{assumption}{Assumption}[section]
\newtheorem{problem}{Problem}[section]
\newtheorem{lemma}{Lemma}[section]

\newtheorem{definition}{Definition}[section]
\newtheorem{theorem}{Theorem}[section]
\newcommand{\qedwhite}{\hfill \ensuremath{\Box}}

{}
\title{Basis transform in  linear switched system models from input-output data}

\author{Fethi~Bencherki, \thanks{Department of Automatic Control, Lund University, Box 117, SE-221 00 Lund, Sweden 
		E-mail: fethi.bencherki@control.lth.se. He was with Eski\c{s}ehir Technical University, Turkey.}  \and Semiha~T\"urkay, 
        $\;\;$ and $\;\;$ H\"{u}seyin Ak\c{c}ay
	    \thanks{Department of Electrical and Electronics Engineering, Eski\c{s}ehir Technical University, 26555 Eskisehir,
	    Turkey, E-mails: \{semihaturkay,huakcay\}@eskisehir.edu.tr. H. Ak\c{c}ay is the {\em corresponding author}. 
	    Tel: +90 222 335 0580 -X 6450.  Fax: +90 222 323 9501. {\em Conflict of interests statement:} This research did not receive any specific grant from funding agencies in the public, commercial, or not-for-profit sectors.} 
    }

\date{December 30, 2021}

\begin{document}

\maketitle

\begin{abstract}
This paper addresses the problem of basis correction in the context of LSS identification from input-output data. It is often the case that 
identification algorithms for the LSSs from input-output data operate locally. The individually identified local submodel estimates reside in 
distinct state bases, which mandates performing a basis correction that facilitates their coherent patching for the ultimate goal of performing 
output predictions for arbitrary inputs and switching sequences. We formulate a persistence of excitation condition for the inputs and the 
switching sequences that guarantee the presented approach's success. These conditions are mild in nature, which proves the practicality of the 
devised algorithm. We supplement the theoretical findings with an elaborating numerical simulation example. 
\end{abstract}


\section{Introduction}

Hybrid systems characterize interactions between discrete and continuous phenomenons. Such systems have seen wide application 
scope recently due to their potential for modeling non-linear time-varying dynamics. They have applications in numerous areas 
of great interest to both scientific and industrial communities, for instance, chemical processes \cite{Lennartsonetal:1996}, 
robotics \cite{Carlonietal:2007,Schlegletal:2003}, air traffic systems \cite{Glover&Lygeros:2004,Prandini&Hu:2008}, networked 
control systems \cite{Daectoetal:2014}, cyber-physical systems \cite{DePersis&Tesi:2015,Cetinkayaetal:2018}, communication 
networks \cite{Leeetal:2007}, power systems \cite{Hiskens&Pai:2000}, human control behavior \cite{Murray-Smith:1998}, computer 
vision \cite{Vidal&Ma:2006}, genetic network modelling  \cite{Cinquemanietal:2008}, and so on.

The LSSs form an important subclass of hybrid systems governed by an external switching sequence. A further division of 
the LSSs as the piecewise linear (PWL) systems and the piecewise affine (PWA) systems is possible. The PWL systems were
introduced by Sontag \cite{Sontag:81} to unify models describing interconnections between automata and linear systems. The PWA 
systems, introduced also in \cite{Sontag:81}, are obtained by partitioning the continuous state and the input vectors into a 
finite number of polyhedral regions where the linear/affine subsystems share the same continuous state in each region. Most 
identification studies, that are available up to now, focus on the input-output hybrid system models \cite{Ferrarietal:2003,Vidal&Soatto&Ma&Sastry:2003,Roll&Bemporad&Ljung:2004,Bemporad&Garulli&Paoletti&Vicino:2005,Bako:2011,
Ozay&Sznaier&Lagoa&Camps:2012,Hojjatinia&Lagoa&Dabbene:2020} even though they are not suitable for the majority of dynamic analysis 
and control methods. Many existing control design methods for hybrid systems are in fact based on state-space models, which 
are also more suitable for dealing with multi-input/multi-output systems. In \cite{Verdult&Verhaegen:2002,Verdult&Verhaegen:2004,
Verdult&Verhaegen:2009}, several subspace algorithms were proposed to identify the PWL and the PWA systems from input-output data 
assuming that the switches are known. Identification of state-space models presents an additional challenge of finding the unknown 
discrete states in comparison to parameters only identification in input-output models. 

The subspace identification algorithms \cite{Verdult&Verhaegen:2002,Verdult&Verhaegen:2004,Verdult&Verhaegen:2009} start by 
estimating local models, called also discrete states or submodels, on time intervals of a sufficient length determined by 
persistence of excitation conditions. A MOESP type algorithm is used to estimate a collection of discrete states up to arbitrary 
similarity transformations. For details on the MOESP algorithms, we refer the reader to \cite{Verhaegen:93,Verhaegen:94} and the 
monograph \cite{Verhaegen&Verdult:2007}. Discrete states identified by a subspace algorithm cannot be combined directly to estimate 
an input-output map of the system since they have been determined only up to similarity transformations. One can select only one 
similarity transformation freely. A critical aspect in every subspace identification algorithm is the observability and the
controllability concepts. Observability plays a central role in realization and identification studies of the LSSs from input-out 
data or the Markov parameters. The observability and the identifiability of the LSSs was studied in \cite{Vidal&Chiuso&Soatto:2002} 
for the homogeneous and in \cite{Elhamifiar:2009} for the inhomogeneous systems. A recent overview of the observability for hybrid 
systems can be found in \cite{DeSantis&DiBenedetto:2016}. The switching times may be estimated from the input-output data by using, 
for example, the change detection method \cite{Pekpe&Mourot&Gasso&Ragot:2004} or the projected subspace classification method
\cite{Borges&Verdult&Verhaegen&Botto:2005}. Iterative identification algorithms were proposed for the LSSs \cite{Borges&Verdult&Verhaegen:2006,Bako&Mercere&Lecoeuche:2009} and the switched auto-regressive with exogenous input (SARX) 
systems \cite{Vidal:2008}. Equivalences of some classes of hybrid systems were investigated in several works
\cite{Heemels&Schutter&Bemporad:2001,Weiland&Juloski&Vet:2006}. In particular, it was shown in \cite{Weiland&Juloski&Vet:2006} 
that any observable PWA system admits a representation as a SARX system and every SARX system 
admits a representation as a PWA system.

Realization theory is an important topic in system theory. It provides a theoretical foundation for model reduction, system 
identification, and observer design. Input-output realization of the PWA models was studied in \cite{Paoletti&Roll&Garulli&Vicino:2007} 
and it was shown that the number of submodels and parameters grows significantly when converting a PWA model into an equivalent 
input-output representation. Realization theory for a special class of the linear parameter varying (LPV) systems and the LSSs was 
studied in \cite{Petretczky&Toth&Mercere:2017} and \cite{Petreczky&Bako&Schuppen:2013}. In the latter work, the minimality of 
the LSS realizations was characterized in terms of the reachability and observability notions. It was shown that the minimal 
realizations are unique up to isomorphisms. Further facts were established in \cite{Petreczky&Bako&Schuppen:2013} as follows. 
Minimality is necessary for identifiability and does not imply minimality of the submodels. Identifiability of an LSS does not 
imply identifiability of submodels. Based on the LSS results in \cite{Petreczky&Bako&Schuppen:2013}, minimality and identifiability 
results for the SARX systems were derived in \cite{Petreczky&Bako&Lecouche&Motchon:2020}. Persistence of excitation notion 
was formulated in \cite{Petreczky&Bako:2023} for the LSSs as a property of hybrid inputs composed of the input and the switching signals. 
It is not tied to any specific algorithm. Another persistence of excitation condition for the SARX systems, which is also not tied 
to any specific algorithm, but to the regressors was derived in \cite{Mu&Chen&Cheng&Bai:2022}. If submodels of an LSS are minimal, 
have a common MacMillan degree $n$, and the switches are separated by a distance of at least $n$, then the LSS turns out to be minimal 
with a dimension $n$. This result established in \cite{Bencherki&Turkay&Akcay:2022} constitutes a converse to
\cite{Petreczky&Bako&Schuppen:2013} under a minimum dwell time constraint. A four-stage algorithm for the realization 
of multi-input/multi-output (MIMO) LSSs from the system Markov parameters was proposed in \cite{Bencherki&Turkay&Akcay:2023} under
various dwell time and system assumptions. An observer-based transformation described in \cite{Bencherki&Turkay&Akcay:2022}
embeds a subset of the LLSs into the set of the SARX systems. This injective mapping allows one to recover submodels of 
a given LLS by sparse optimization up to arbitrary similarity transformations. Unless there is only one submodel, the LLS model
delivered by this scheme will not be isomorphic to the targeted LSS. We refer the interested reader to the recent survey 
\cite{Bako:2023} for applications of the sparsity-inducing methods to system identification. 

\subsection{Organization of the paper}
In Section~\ref{basistransform}, we formulate the basis transformation problem and motivate its importance in hybrid system identification. 
We survey related works in the literature and highlight our contributions. In Section~\ref{mainsec}, we derive interpolation conditions for 
basis transformation algorithms which allow one to uniquely extract products of similarity transformations from the input-output data 
assuming that certain rank conditions are met. This part provides a solution to the problems studied in \cite{Vidal&Chiuso&Soatto:2002,Verdult&Verhaegen:2002,Verdult&Verhaegen:2004,Verdult&Verhaegen:2009} given an input-output data set 
and a switching signal. Reproducibility of this solution for new data sets depends on the persistence of excitation conditions for 
the subclass of the LLSs under consideration. We design persistently exciting (PE) hybrid inputs in Section~\ref{hybridinpsec} for a 
subclass of the LSSs used in most subspace identification studies of hybrid systems. Relations of the basis transformation problem 
with graph theory are explored in Section~\ref{graphsec}. A basis transforming algorithm and the main result of the paper are presented 
in Section~\ref{secbasistf}. Section~\ref{numinsec} is devoted to a numerical study of the proposed algorithm. Section~\ref{consec} 
concludes the paper.

\section{Problem formulation}~\label{basistransform}
In this section, we will formulate the basis transformation problem studied in this paper. Let
\begin{eqnarray}
	x(k+1) &=& A_{\varphi(k)} x(k) +B_{\varphi(k)} u(k) \label{ssx} \\
	y(k) &=& C_{\varphi(k)} x(k) + D_{\varphi(k)} u(k) \label{ssy}
\end{eqnarray}
be the true LSS where $u(k) \in \mathbb{R}^m$, $y(k) \in \mathbb{R}^p$, $x(k) \in \mathbb{R}^{n}$ are respectively the continuous 
input, the output, and the continuous-state, $\varphi(k):[1\;\;N] \mapsto \mathbb{S}=\left\{1,\cdots, \sigma \right\}$ is the switching 
sequence for fixed $\sigma$ and $N$, and ${\mathcal P}_{\varphi(k)}=(A_{\varphi(k)},B_{\varphi(k)},C_{\varphi(k)},D_{\varphi(k)})$,
$0\leq k \leq N$ are the discrete states. Let ${\mathcal P}=\{{\mathcal P}_1,\cdots,{\mathcal P}_\sigma\}$ denote the discrete state set. We assume that $\varphi$ is a surjective function, i.e., it satisfies ${\rm range}(\varphi)=\mathbb{S}$. As in the works \cite{Petreczky&Bako&Schuppen:2013,Petreczky&Bako:2023,Petreczky&Bako&Lecouche&Motchon:2020}, we view $\varphi$ as an external 
input and define the hybrid inputs $\omega$ by
\begin{equation}\label{hybrid-input}
\omega=((\varphi(0),u(0)), \cdots, (\varphi(N-1),u(N-1))).
\end{equation}
Thus, $y(k)$, $0\leq k\leq N$ is the response of the LSS in (\ref{ssx})--(\ref{ssy}) to $\omega$ while respecting the causality. In 
(\ref{ssx})--(\ref{ssy}), ${\mathcal P}$ is fixed and $\varphi$ and $u$ are changed. We tacitly assume that $\omega$ is the $N$-truncation 
of an infinite string. To emphasize the dependence on ${\mathcal P}$, we write $y_{\mathcal P}(\omega)$ for the output sequence produced 
by $\omega$. The basis transformation algorithms in this paper do not need to know $x(0)$ since they implicitly estimate it.

Recall that $\varphi$ partitions $\left[1\;\;N\right]$ into disjoint segments $[k_i,k_{i+1})$ such that $\varphi(k) = \varphi(k_i)$ 
for all $k$ in the segment $[k_i \;\;k_{i+1})$ with $k_0=1<k_1 < \cdots < k_{i^*} <N=k_{i^*+1}=N$ where $k_0$ and $k_{i^*+1}$ have been 
introduced for convenience, not to be confused with the switches of (\ref{ssx})--(\ref{ssy}) in the set $\chi=\{k_i: 1\leq i \leq i^*\}$. 
The dwell times 
and the minimum dwell time of (\ref{ssx})--(\ref{ssy}) are defined respectively by $\delta_i(\chi) = k_{i+1}-k_i$ for 
$0 \leq i \leq i^*$ and $\delta_*(\chi)=\min_{0\leq i \leq i^*}\delta_i(\chi)$. We say two linear time-invariant (LTI) systems 
${\mathcal W}_1=(A_1,B_1,C_1,D_1)$ and ${\mathcal W}_2=(A_2,B_2,C_2,D_2)$ are similar and show this relation by 
${\mathcal W}_2 \sim {\mathcal W}_1$ if there  exists a nonsingular matrix $T$ satisfying the equations $A_2=T^{-1} A_1 T$, $C_2=C_1 T$, 
$B_2=T^{-1} B_1$, and $D_2=D_1$. Recall that transfer functions and Markov parameters of two similar systems are equal. The requirements 
on the LSS are captured in the  following. These requirements pertain to a proper subclass of the LSSs.

\begin{assumption}[Model set]\label{stabmin}
The discrete states ${\mathcal P}_j \in {\mathcal P}$, $1\leq j \leq \sigma$ have a common Macmillan degree $n$, no poles at zero,
and they are bounded-input/bounded-output (BIBO) stable. 
\end{assumption}

Assumption~\ref{stabmin} implies that (\ref{ssx})--(\ref{ssy}) is {\em span reachable} and {\em observable}, thus {\em minimal} 
while the converse is not necessarily true \cite{Petreczky&Bako&Schuppen:2013}. It is also a well-known fact that if two minimal 
LSSs are equivalent, that is, if they are the minimal realization of the same input-out map, then they are isomorphic. 
Assumption~\ref{stabmin} is a natural one since it permits recovery of local models from input-output data by an identification 
algorithm. For example, a subspace algorithm returns submodel estimates in minimal state-space realizations. To understand what 
happens when the discrete states are estimated locally, let $\hat{\mathcal P}_\nu$, $1\leq \nu \leq \sigma$ be a collection of 
discrete state estimates similar to the discrete states in ${\mathcal P}$, that is, 
\begin{equation}\label{similar}
		\hat{A}_\nu=T_\nu^{-1} A_\nu T_\nu, \;\;\; \hat{B}_\nu=T_\nu^{-1} B_\nu,\;\;\;
        \hat{C}_\nu=C_\nu T_\nu,\;\;\;\hat{D}_\nu=D_\nu.
\end{equation}
for some nonsingular matrices $T_\nu$, $1\leq \nu \leq \sigma$. The isomorphism would require $T_\nu = T$ for all $\nu$ and some 
nonsingular $T$. It would even require the relation $\hat{x}(0)=T^{-1}x(0)$ to hold between the initial states \cite{Petreczky&Bako&Schuppen:2013}. The isomorphism mandates a basis transform on local models as described in this paper. 

Another issue with the local models is that if $\varphi$ is not injective, some of the following clusters 
$$
\Pi_\nu=\{i: \,\varphi(k_i)=\nu,\;0 \leq i \leq i^*\}, \qquad \nu=1,\cdots,\sigma
$$
will not be singleton. For such a $\nu$, the discrete state estimates in the segments $[k_i\;\;k_{i+1})$, $i \in \Pi_\nu$ may be 
in different similarity transformations per segment; but, they will all be similar to ${\mathcal P}_\nu$ if they recover it up to 
arbitrary similarity transformations in each segment. In this case, we can pick a representative discrete state estimate 
$\hat{\mathcal P}_\nu$ such that $\hat{\mathcal P}_\nu \sim {\mathcal P}_\nu$, $1\leq \nu \leq \sigma$. A clustering algorithm 
\cite{Esteretal:96} may be used to identify the sets $\Pi_v$, $1\leq \nu \leq \sigma$ under the following condition on ${\mathcal P}$.

\begin{assumption}[Distinguishability]\label{unimodular}
The discrete states ${\mathcal P}_j$, $1\leq j \leq \sigma$ satisfy
\begin{equation}\label{statistic0}
{\mathcal M}(A_{\varphi(k_i)}) \neq {\mathcal M}(A_{\varphi(k_j)}) \Longleftrightarrow \varphi(k_i) \neq \varphi(k_j)
\end{equation}
where ${\mathcal M}(X)$ is the feature used for clustering defined as the $\ell_1$ norm of the eigenvalues
\begin{equation}\label{statistic}
{\mathcal M}(X)=\sum_{j=1}^n |\lambda_j(X)|, \qquad X \in \mathbb{R}^{n \times n}.
\end{equation}
\end{assumption}
Recall that $\varphi$ is surjective. This means that the cardinality of $\Pi_\nu$ denoted by $|\Pi_\nu|$, i.e., the number of 
the elements in $\Pi_\nu$ is not zero for all $\nu$. If  for all $0\leq i \leq i^*$, the discrete state estimates on $[k_i\;\;k_{i+1})$ 
are similar to $\mathcal{P}_{\varphi(k_i)}$, under Assumption~\ref{unimodular} we can extract one discrete state estimate from each cluster
with the property $\hat{\mathcal P}_\nu \sim \mathcal {P}_\nu$, $1 \leq \nu \leq \sigma$. We will state this as a standing assumption
on the local estimates.

\begin{assumption}[Local recovery]\label{locrec}
For all $i \in [0\;\;i^*]$, the local estimate on $[k_i\;\;k_{i+1})]$ is similar to $\mathcal{P}_{\varphi(k_i)}$.
\end{assumption}

Let $\hat{\mathcal P}=\{\hat{\mathcal P}_\nu:\, 1 \leq \nu \leq \sigma\}$ denote this local model set obtained by a clustering algorithm. 
We impose a mild condition on the minimum dwell time as follows.

\begin{assumption}[Minimum dwell time]\label{mindwell}
The switching sequence $\varphi$ satisfies $\delta_*(\chi) \geq n$. 
\end{assumption}
Assumption~\ref{mindwell} will be needed to learn the input-output map $y_{\mathcal P}$. Once this map is learned, $y_{\mathcal P}(\omega)$ 
can be  calculated for any hybrid input $\omega$, without fulfilling Assumption~\ref{mindwell}, as shown in the article. 

The discrete states $\hat{\mathcal P}_\nu$ are related to $\mathcal{P}_\nu$ via the fixed similarity transformation $T_\nu$ as a result of 
the selection procedure. Since both ${\mathcal P}_\nu$ and $T_\nu$ are unknown, there is no direct way of estimating the two from 
(\ref{similar}) only. Instead, as we shall see one way is to match the outputs $y_{\mathcal P}(\omega)$ and 
$y_{\hat{\mathcal P}}(\omega)$ by finding the products of the similarity transformations. Therefore, for $1 \leq \nu \leq \sigma$ we replace 
(\ref{similar}) with
\begin{equation}\label{similar2}
		\check{A}_\nu=\check{T}_\nu^{-1} \hat{A}_\nu \check{T}_\nu, \;\;\; \check{B}_\nu=\check{T}_\nu^{-1} \hat{B}_\nu,\;\;\;
        \check{C}_\nu=\hat{C}_\nu \check{T}_\nu,\;\;\;\check{D}_\nu=\hat{D}_\nu
\end{equation}
and set $\check{\mathcal P}_\nu=(\check{A}_\nu,\check{B}_\nu,\check{C}_\nu,\check{D}_\nu)$ and 
$\check{\mathcal P}=\{\check{\mathcal P}_\nu:\, 1 \leq \nu \leq \sigma\}$. The objective is the enforcement of the output matching  
$y_{\mathcal P}(\omega)=y_{\check{\mathcal P}}(\omega)$ by suitably selecting $\check{T}_\nu$ in (\ref{similar2}) for a given or all $\omega$. 
Since the similarity relation is transitive, note that $\check{\mathcal P}_\nu \sim {\mathcal P}_\nu$ for all $1\leq \nu \leq \sigma$. 
We state our first objective as follows.

\begin{problem}\cite{Verdult&Verhaegen:2004} \label{probform1}
    Suppose that Assumptions~\ref{stabmin}--\ref{mindwell} hold. Given $(\omega,\hat{\mathcal P})$, find $\check{T}_\nu$ in (\ref{similar2}) 
    for $1\leq \nu \leq \sigma$ such that $y_{\mathcal P}(\omega)=y_{\check{\mathcal P}}(\omega)$.
\end{problem} 

Problem~\ref{probform1} is a restatement of the state-basis transformation problem formulated in \cite{Verdult&Verhaegen:2004} borrowing 
the terminology of \cite{Petreczky&Bako:2023}. Several questions come to mind. The first one is if $y_{\hat{\mathcal P}}(\omega)$ 
is well-defined. Recall that $\hat{\mathcal P}$ was obtained by fixing $\omega$ and for each $1 \leq \nu \leq \sigma$ selecting one 
discrete state estimate on $[k_i\;\;k_{i+1})$ for some $i \in \Pi_\nu$. Conceivably  another choice $\tilde{\mathcal P}$ would yield 
$y_{\tilde{\mathcal P}}(\omega) \neq y_{\hat{\mathcal P}}(\omega)$. The second question is related to reproducibility. More specifically, 
suppose given $\omega$, $\check{T}_\nu$ were found to enforce $y_{\check{\mathcal P}}(\omega)=y_{\mathcal{P}}(\omega)$ and now a new 
switching signal $\varphi^\prime \neq \varphi$ is applied to (\ref{ssx})--(\ref{ssy}) with the same submodel set $\check{\mathcal P}$. 
It is legitimate to ask if $y_{\mathcal P}(\omega^\prime) =y_{\check{\mathcal P}}(\omega^\prime)$ where 
$\omega^\prime =((\varphi^\prime(0),u^\prime(0)),\cdots,(\varphi^\prime(N-1),u^\prime(N-1)))$. Note that both $u=u^\prime$ and
$u \neq u^\prime$ cases are possible. It would be tedious to recalculate $\check{T}_\nu$ every time a new switching sequence 
$\varphi^\prime \neq \varphi$ is presented. A third question is if it is possible to drop the restriction $\delta_*(\chi) \geq n$, 
once $\check{\mathcal P}$ has been learned, that is, $y_{\mathcal P}(\omega^\prime) = y_{\check{\mathcal P}}(\omega^\prime)$ for all
$\omega^\prime$, in another words, $y_{\check{\mathcal P}}=y_{\mathcal P}$. Positive answers to these queries place restrictions 
on hybrid inputs. Designing hybrid inputs for learning $\check{\mathcal P}$ with the property $y_{\check{\mathcal P}}=y_{\mathcal P}$, 
in particular the ones with shortest length is expected to impact identifiability and parametrization of hybrid systems.
The main problem extends Problem~\ref{probform1} to arbitrary $\omega$ as follows.

\begin{problem}[Main problem]\label{probform2}
    Suppose that Assumptions~\ref{stabmin}--\ref{mindwell} hold. Given $(\omega,\hat{\mathcal P})$, find $\check{T}_\nu$ in (\ref{similar2})
    for $1\leq \nu \leq \sigma$  such that $y_{\mathcal P}=y_{\check{\mathcal P}}$.
\end{problem}
   
\subsection{Motivation for a basis transformation}

When $\chi$ is known, the MOESP algorithms \cite{Verhaegen:93,Verhaegen:94} can be used to retrieve the discrete states from the 
nput-output data up to similarity transformations assuming that the continuous inputs are PE \cite{Verdult&Verhaegen:2004} over the 
segments of interest. The discrete-state estimates, however, cannot be combined directly 
because each discrete state lies in a different state basis. Methods for detecting switches were proposed in 
\cite{Pekpe&Mourot&Gasso&Ragot:2004,Borges&Verdult&Verhaegen&Botto:2005,Borges&Verdult&Verhaegen:2006}.  A common feature of them 
is the computation of similarity transformations to bring the discrete-state estimates to the same basis. The order of estimation was 
changed in the more recent works \cite{Bencherki&Turkay&Akcay:2022,Bencherki&Turkay&Akcay:2023}. First, the discrete states were 
estimated by minimizing a sparsity promoting norm. Then, the switches were estimated using a MOESP type subspace algorithm from the 
input-out data. In these works as well, a basis transformation stage complements the identification procedure. A feasible solution to 
Problem~\ref{probform2} requires that (\ref{ssx})--(\ref{ssy}) is identifiable and $\omega$ is PE. 
In \cite{Petreczky&Bako:2023}, see Theorem~5, asymptotic identifiability results  were derived, but identifiability and accuracy
issues for the LSSs excited by hybrid inputs of finite length were not addressed, see Remark~19 in the same paper. The 
second goal of this paper is to show that there are overwhelmingly abundant PE inputs of length $N$ satisfying $N \leq c \sigma n$ for 
some constant $c>0$ that does not depend on $N$ for the subclass of the LSSs in Assumption~\ref{stabmin}. 

\subsection{Related work}  

The basis transformation issue in the LSS models may be circumvented by resorting to canonical forms. Transformation of an LTI 
state-space model to the observable canonical form was previously studied in \cite{Kailath:1980} for single-input/single-output 
(SISO) or multi-input/single-output (MISO) systems. For this approach to be applicable, the original system must be in a canonical 
form. Therefore, it suffices to transform the identified discrete states to canonical forms. For an observable MIMO system with 
characteristic polynomial equal to its minimal polynomial,  that is, a system with a non-derogatory state-transition matrix, an 
observability canonical form can be derived by following a randomization procedure in \cite{Mercere&Bako:2011}. Closely related 
works to some parts of our work  are \cite{Verdult&Verhaegen:2004,Borges&Verdult&Verhaegen:2006}. This paper extends the basis 
transformation results derived in these works as outlined in Section~\ref{subseccontrib}. A solution to the basis transformation 
problem was presented in \cite{Bencherki&Turkay&Akcay:2023} for the special case of pulse continuous inputs acting on the LSS in 
(\ref{ssx})--(\ref{ssy}).    

\subsection{Contributions}\label{subseccontrib}

We solve Problem~\ref{probform2} and compare our contributions to the existing works in the literature:

\begin{enumerate}

	\item Problem~\ref{probform1} was considered in \cite{Vidal&Chiuso&Soatto:2002} for autonomous systems using only one transition 
    from one discrete state to another to determine a state transformation. As pointed out in \cite{Verdult&Verhaegen:2004} this leads 
    to an underdetermined set of equations for which there are many solutions which are all compatible with data used for identification, 
    but not necessarily with another data set. This paper, similarly to \cite{Verdult&Verhaegen:2004} on the other hand, uses all 
    available transitions to uniquely determine transformations.
    
	\item  In \cite{Verdult&Verhaegen:2004}, a state-basis transformation was determined by relating states before and after a jump.
	Since the continuous state vector evolves according to (\ref{ssx}), the method proposed in \cite{Verdult&Verhaegen:2004} is not 
    applicable to the input-output map setting considered in this paper. The input-out problem formulation, on the other hand, generalizes 
    the setup in \cite{Verdult&Verhaegen:2004} by letting $y(k)=x(k)$. 
	
	\item The method in \cite{Verdult&Verhaegen:2004} is based on the principle of uniquely determining a matrix relating
    continuous states before and after a jump assuming that a sufficient number of transitions takes place between the same pair 
    of submodels. These matrices are not related to the transfer functions of the submodels. In this paper, we first choose 
    submodel estimates in one-to-one correspondence with the true submodels and match the true and the estimated Markov parameters 
    using the input-output data. PE hybrid inputs are the only things needed; but, we show that plenty of them are available.
	
	\item The PE concept for the LSSs introduced in the literature are intertwined with the basis transformation problem in this  paper. 
     Enlargement of continuous inputs to hybrid inputs yields a neat solution to Problem~\ref{probform2}. A procedure to construct 
     a basis transformation is presented. As a byproduct, existence of PE inputs for the subclass in Assumption~\ref{stabmin} 
     satisfying $N=O(\sigma n)$ is shown. Here, the notation $y=O(x)$ means that $|y| \leq c |x|$ for some 
     $c>0$ that does not depend on $x$. 
	
	\item A link between the basis transformation problem and the graph theory is put forward. A numerical example demonstrates
	effectiveness of the presented basis construction approach.
	  
\end{enumerate} 

\section{Interpolation conditions for basis transformation}\label{mainsec}

Given $1 \leq i \leq i^*$, from the observability of ${\mathcal P}_{\nu}$ and ${\mathcal P}_{\mu}$ where $\nu=\varphi(k_{i-1})$ 
and $\mu=\varphi(k_i)$, first we estimate $x(t_{i-1})$, $t_{i-1} \in [k_{i-1}\;\;k_i-2]$ from the input-output data 
$(u(t_{i-1}),y(t_{i-1})),\cdots,(u(\eta_{i-1}),y(\eta_{i-1}))$ where $t_{i-1} < \eta_{i-1} \leq  k_i-1$. Let $q_{i-1}=\eta_{i-1}-t_{i-1}$.  
Recursively using (\ref{ssx})--(\ref{ssy}), we derive
\begin{equation}\label{Ykl}
	Y_{i-1} = {\mathcal O}_{i-1} x(t_{i-1})+\Gamma_{i-1} U_{i-1}
\end{equation}
where
\begin{eqnarray}
Y_{i-1}&=& [y^T(t_{i-1}) \;\cdots \;y^T(\eta_{i-1})]^T, \nonumber \\
U_{i-1} &=& [u^T(t_{i-1}) \;\cdots \;u^T(\eta_{i-1})]^T, \nonumber \\
{\mathcal O}_{i-1} &=& \left[\begin{array}{c} C_\nu \\ \vdots \\ C_\nu 
A^{q_{i-1}}_\nu \end{array} \right], \label{Obser}\\ 
\Gamma_{i-1} &=& \left[\begin{array}{ccc}  D_\nu  &  & 0 \\ \vdots & \ddots & \vdots \\ 
	  C_\nu  A^{q_{i-1}-1}_\nu B_\nu & \cdots &  D_\nu 
   \end{array} \right]. \label{Gamma} 
\end{eqnarray}
Recall that $\hat{A}_\nu=T^{-1}_\nu A_\nu T_\nu$, $\hat{C}_\nu=C_\nu T_\nu$, $\hat{B}_\nu=T^{-1}_\nu B_\nu$, and 
$\hat{D}_\nu=D_\nu$. Let $\hat{\mathcal O}_\nu$ and $\hat{\Gamma}_\nu$ denote the extended observability and the lower 
triangular block Toeplitz matrices calculated with $\hat{A}_\nu$, $\hat{B}_\nu$, $\hat{C}_\nu$, and $\hat{D}_\nu$. Note that 
$\hat{\mathcal O}_\nu={\mathcal O}_\nu T_\nu$ and $\hat{\Gamma}_\nu={\Gamma}_\nu$. Assume $q_{i-1} \geq n-1$. We estimate
$x(t_{i-1})$ from (\ref{Ykl}) as
\begin{equation}\label{hatxk}
\hat{x}(t_{i-1})=\hat{\mathcal O}_{i-1}^\dag \left(Y_{i-1}-\hat{\Gamma}_{i-1}U_{i-1}\right) 
\end{equation}
where we used $\hat{\Gamma}_\nu={\Gamma}_\nu$ and $X^\dag=(X^TX)^{-1}X^T$ denotes the Moore-Penrose pseudo inverse of a given full 
column-rank matrix $X$. Substituting the similarity equations above in (\ref{hatxk}), note that
\begin{equation}\label{surahihanim}
\hat{x}(t_{i-1})=T_\nu^{-1} x(t_{i-1}). 
\end{equation}
 
Since $q_{i-1}=\eta_{i-1}-t_{i-1}>0$ and $q_{i-1} \geq n-1$ by assumption, we select $n\geq 2$ which makes (\ref{hatxk}) valid. 
This shall be assumed in the sequel. The case $n=1$ is ignored since it is uninteresting. Because $t_{i-1}=k_{i-1}$ and $\eta_{i-1}=k_i-1$ are 
feasible and $q_{i-1} \geq n-1$ by assumption, we derive the natural dwell time condition $\delta_{i-1}(\chi) \geq n$ as presented in Assumption~\ref{mindwell}. Notice that $k_{i-1} \leq t_{i-1}=\eta_{i-1}-q_{i-1} \leq k_i-n$. 

Now, we calculate the output $y(k)$ based on the inputs $u(t)$, $t_{i-1} \leq t \leq k$ and the initial state $x(t_{i-1})$. 
Recursively using (\ref{ssx})--(\ref{ssy}), we derive for all $k \in [k_i\;\;k_{i+1})$ 
\begin{equation}\label{summand}
y(k)=C_{\varphi(k)} \Phi(k,t_{i-1}) x(t_{i-1})+\sum_{l=t_{i-1}}^k h(k,l)u(l)
\end{equation}
where the Markov parameters and the state transition matrix are defined for (\ref{ssx})--(\ref{ssy}) 
by
\begin{eqnarray}
	h(k,k)= D_{\varphi(k)} \;\;\;&\mbox{and}&\;\;\; h(k,l) = C_{\varphi(k)}\Phi(k,l+1) B_{\varphi(l)}, \;\; l<k \nonumber 
	\\[-1.5ex] \label{hPhidef}  \\[-1.5ex]
	\Phi(k,k) = I_n\;\;&\mbox{and}&\;\; \Phi(k,l+1) = A_{\varphi(k-1)}\;\cdots\;A_{\varphi(l+1)},\;\;l <k-1. \nonumber
\end{eqnarray}
Consider the $h(k,l)$ terms in (\ref{summand}). If $k_i \leq l \leq k-2$, from (\ref{hPhidef})
\begin{eqnarray}
h(k,l) &=& C_{\varphi(k)}A_{\varphi(k-1)} \;\cdots \;A_{\varphi(l+1)}B_{\varphi(l)} \nonumber 
\\[-1.5ex] \label{cukka21} \\[-1.5ex]
&=& C_\mu A^{k-l-1}_\mu B_\mu = \hat{C}_\mu \hat{A}^{k-l-1}_\mu \hat{B}_\mu \nonumber
\end{eqnarray}
where we used the similarity equations above not only for $\nu$, but also for $\mu$. If $k_i \leq l = k-1$, from the right-hand side of 
the first equation in (\ref{hPhidef}), $h(k,l)=\hat{C}_\mu\hat{B}_\mu$. It follows that (\ref{cukka21}) holds for all $k_i \leq l \leq k-1$. 
Note also that $h(k,k)=\hat{D}_\mu$ for all $k \in [k_i\;\;k_{i+1})$. Now, if $l \leq k_i-2$ and $k>k_i$, from (\ref{hPhidef})
\begin{eqnarray}
	h(k,l) &=& C_{\varphi(k)}A_{\varphi(k-1)}\;\cdots\; A_{\varphi(k_i)}A_{\varphi(k_i-1)} \;\cdots \;A_{\varphi(l+1)}B_{\varphi(l)} \nonumber 
	\\[-0.5ex] \label{cukka22}\\[-1ex]
	&=& C_\mu A^{k-k_i}_\mu A^{k_i-l-1}_\nu B_\nu= \hat{C}_\mu \hat{A}^{k-k_i}_\mu T_\mu^{-1} T_\nu
    \hat{A}^{k_i-l-1}_\nu\hat{B}_\nu. \nonumber 	
\end{eqnarray}
If $k=k_i \geq l+2$, then
\begin{eqnarray*}
	h(k,l) &=& C_{\varphi(k)}A_{\varphi(k-1)}\;\cdots \; A_{\varphi(l+1)}B_{\varphi(l)} =  C_\mu A^{k_i-l-1}_\nu B_\nu \nonumber 
	\\[-0.5ex] \label{cukka222}\\[-1ex]
	&=&  \hat{C}_\mu T_\mu^{-1} T_\nu \hat{A}^{k_i-l-1}_\nu \hat{B}_\nu \nonumber 	
\end{eqnarray*} 
which is (\ref{cukka22}) evaluated at $k=k_i$. Thus, the last equation in (\ref{cukka22}) holds for all $l+2 \leq k_i \leq k$. 
When $l=k_i-1$, there are two cases. If $k=k_i$, then $h(k,l)=C_\mu B_\nu = \hat{C}_\mu T_\mu^{-1} T_\nu \hat{B}_\nu$. 
Similarly, if $k>k_i$ we get $h(k,l) = \hat{C}_\mu \hat{A}^{k-k_i}_\mu T_\mu^{-1} T_\nu \hat{B}_\nu$. 
Hence, the last equation in (\ref{cukka22}) holds for all $l+1 \leq k_i \leq k$. Next, from $x(t_{i-1})=T_\nu \hat{x}(t_{i-1})$ 
in (\ref{surahihanim}) and (\ref{hPhidef}) for all $k \geq k_i$,
\begin{eqnarray}
C_{\varphi(k)} \Phi(k,t_{i-1}) x(t_{i-1}) &=& C_\mu A^{k-k_i}_\mu A^{k_i-\xi_{i-1}}_\nu x(t_{i-1}) \nonumber 
\\[-.5ex] \label{cukka23} \\[-.5ex] 
&=& \hat{C}_\mu \hat{A}^{k-k_i}_\mu T_\mu^{-1} T_\nu \hat{A}^{k_i-t_{i-1}}_\nu \hat{x}(t_{i-1}). 
\nonumber
\end{eqnarray}
Hence, from (\ref{summand}) and (\ref{cukka21})--(\ref{cukka23}), 
\begin{equation}\label{zk}
\zeta_i(k) =\hat{C}_\mu \hat{A}^{k-k_i}_\mu T_\mu^{-1} T_\nu \kappa_{i-1}(t_{i-1}), \qquad  \forall k \in [k_i\;\;k_{i+1})
\end{equation}
where
\begin{eqnarray}
\zeta_i(k) &\stackrel{\Delta}{=} & \left\{\begin{array}{lr} y(k) -\hat{D}_\mu u(k)-\displaystyle{\sum_{l=k_i}^{k-1} 
\hat{C}_\mu \hat{A}^{k-l-1}_\mu \hat{B}_\mu u(l)}, &k>k_i \\ y(k) -\hat{D}_\mu u(k), & k=k_i 
\end{array} \right. \label{zkdef1} \\
\kappa_{i-1}(t_{i-1}) &=& \hat{A}^{k_i-t_{i-1}}_{\varphi(k_{i-1})} \hat{x}(t_{i-1})+\sum_{l=t_{i-1}}^{k_i-1} 
\hat{A}^{k_i-l-1}_\nu \hat{B}_\nu u(l). \label{kapd}
\end{eqnarray}
Given $i \in [1\;\;i^*+1]$, for each pair $(k,t_{i-1})$ one first computes $\zeta_i(k)$ and $\kappa_{i-1}(t_{i-1})$ from 
(\ref{zkdef1})--(\ref{kapd}) using the input-output data $((u(t_{i-1}),y(t_{i-1})),\cdots,(u(k),y(k)))$, $\hat{x}(t_{i-1})$ in 
(\ref{hatxk}), and the state-space matrices of $\hat{\mathcal P}_\nu$ and $\hat{\mathcal P}_\mu$. Then, (\ref{zk}) becomes 
a linear equation in $T_\mu^{-1} T_\nu$. A question arises if any new information is gained as $t_{i-1}$ and $k$ are changed within 
their limit values. The intuitive answer for $t_{i-1}$ is negative since this change produces only redundant information. A proof is however necessary and provided next.  

\begin{lemma}\label{lem1}
	Suppose that $n \geq 2$, $i\geq 1$, and Assumptions~\ref{stabmin}--\ref{mindwell} hold. Let $\kappa_{i-1}(t_{i-1})$ be as 
     in (\ref{kapd}). Then, $\kappa_{i-1}(t_{i-1})=\kappa_{i-1}(k_i-n)$ for all $t_{i-1} \in [k_{i-1}\;k_i-n]$.
\end{lemma}

{\em Proof.} Starting from $k=t_{i-1}$ write (\ref{ssx}) backward in time
\begin{equation}\label{fgh}
x(t_{i-1}-1)=A^{-1}_{\varphi(k_{i-1})} x(t_{i-1}) - A^{-1}_{\varphi(k_{i-1})} B_{\varphi(k_{i-1})} u(t_{i-1}-1).
\end{equation}
Recall that $\hat{x}(\xi_{i-1})=T_v^{-1} x(t_{i-1})$ for all $k_{i-1} \leq t_{i-1} \leq k_i-n$ where $\nu=\varphi(k_{i-1})$. The 
derivation of this equation relies on the observability of the discrete states.  Then, from (\ref{fgh}) if $t_{i-1}>k_{i-1}$,
\begin{eqnarray}
\hat{x}(t_{i-1}-1)&=& T_\nu^{-1} x(t_{i-1}-1) \nonumber\\
&=& T_v^{-1} A^{-1}_\nu x(t_{i-1})-T_v^{-1} A^{-1}_\nu B_\nu \, u(t_{i-1}-1) \nonumber\\
&=& \hat{A}^{-1}_\nu T_\nu^{-1} x(t_{i-1})-\hat{A}^{-1}_\nu \hat{B}_\nu \, u(t_{i-1}-1) \nonumber\\
&=& \hat{A}^{-1}_\nu \hat{x}(t_{i-1})-\hat{A}^{-1}_\nu \hat{B}_\nu \, u(t_{i-1}-1). \label{mantarsin}
\end{eqnarray}
Since we assumed $t_{i-1}>k_{i-1}$ above, $t_{i-1}-1 \geq k_{i-1}$ and $t_{i-1}$ is within limits. Plug $t_{i-1}-1$ in (\ref{kapd})
\begin{eqnarray*}
\kappa_{i-1}(t_{i-1}-1)&=&\hat{A}^{k_i-t_{i-1}+1}_\nu \hat{x}(t_{i-1}-1)+\sum_{l=t_{i-1}-1}^{k_i-1} 
\hat{A}^{k_i-l-1}_\nu \hat{B}_\nu \, u(l) \\
&=&  \hat{A}^{k_i-t_{i-1}}_\nu \hat{x}(t_{i-1}) - \hat{A}^{k_i-t_{i-1}}_\nu \hat{B}_\nu \,  u(t_{i-1}-1) \\
&{}& \hspace{-5mm} +\sum_{l=t_{i-1}}^{k_i-1}\hat{A}^{k_i-l-1}_\nu
\hat{B}_\nu \, u(l) + \hat{A}^{k_i-t_{i-1}}_\nu \hat{B}_\nu \, u(t_{i-1}-1) \\
&=& \kappa_{i-1}(t_{i-1})	
\end{eqnarray*}
where the middle equality followed from (\ref{mantarsin}). The proof carries on by recursion and the iterations stop at $t_{i-1}=k_{i-1}$. 
$\qedwhite$

We fix $t_{i-1}=k_{i-1}$ and $q_{i-1}=\delta_*(\chi)-1$ for all $i \in [1\;\;i^*+1]$. Evaluate $\zeta_i(k)$ for $k \in [k_i \;\;q_i]$ 
and stack the resulting vectors into a compound vector $Z_i=[\zeta_i^T(k_i)\;\cdots\;\zeta_i^T(k_i+q_i)]^T$. Define 
$\Upsilon:\mathbb{S}^2 \mapsto \mathbb{R}^{n \times n}$ by
\begin{equation}\label{Upsilondef}
	\Upsilon(\nu,\mu)=T_\mu^{-1} T_\nu.
\end{equation}
Since $q_i \geq n-1$, from the Cayley-Hamilton theorem notice that all the information in $\zeta_i(k)$, $k_i \leq k <k_{i+1}$ 
is extracted by restricting $k$ to $[k_i \;\;q_i]$. Although $T_\mu^{-1}$ and $T_\nu$ are unique, we haven't made any claim if 
they or their product can be determined from the input-output data. This requires further work. The hybrid inputs need to be PE 
s we shall see next. Given the pairs $(\nu,\mu) \in \mathbb{S}^2$, we define the ordered sets
\begin{equation}
	{\mathcal N}_{\nu,\mu} = \left\{i:\, \varphi(k_{i-1})=\nu,\; \varphi(k_i)=\mu \right\}. \label{chij1j2}
\end{equation}
Note that ${\mathcal N}_{\nu,\nu}$ is empty on $\mathbb{S}$ and $\hat{\mathcal O}_i$ is constant on ${\mathcal N}_{\nu,\mu}$. 
From (\ref{zk}), 
\begin{equation}\label{Ziequation}
	Z_i = \hat{\mathcal O}_i  \Upsilon(\nu,\mu) \kappa_i(k_{i-1}), \qquad i \in {\mathcal N}_{\nu\mu}.
\end{equation}
By stacking $Z_i$ and $\kappa_{i-1}(k_{i-1})$ along columns as $i$ assumes increasing values in ${\mathcal N}_{\nu,\mu}$, we define
two compound matrices ${\mathcal Z}_{\nu,\mu}$ and $\Psi_{\nu,\mu}$. Pick the smallest element in ${\mathcal N}_{\nu,\mu}$ and 
denote it by $i_{\nu,\mu}$. Then, from (\ref{Ziequation}) we get 
${\mathcal Z}_{\nu,\mu} = \hat{\mathcal O}_{i_{\nu,\mu}} \Upsilon(\nu,\mu) \Psi_{\nu,\mu}$.
From the observability of the discrete states, we derive
\begin{equation}\label{Zcomp}
\hat{\mathcal O}_{i_{\nu,\mu}}^\dag  {\mathcal Z}_{\nu,\mu}=\Upsilon(\nu,\mu) \Psi_{\nu,\mu}.
\end{equation}
This equation furnishes all available information to solve Problem~\ref{probform1} since it was derived using the entire input-output 
data set. By changing $(\nu,\mu)$ over non-empty ${\mathcal N}_{\nu,\mu}$ sets, $\Upsilon(\nu,\mu)$ is determined either uniquely or 
non-uniquely. If $\Psi_{\nu,\mu}$ has full-row rank, the unique solution of (\ref{Zcomp}) is obtained as follows.  
\begin{equation}\label{Upsilonnumu}
\Upsilon(\nu,\mu) = \hat{\mathcal O}_{i_{\nu\mu}}^\dag  {\mathcal Z}_{\nu,\mu} \Psi_{\nu,\mu}^\dag.
\end{equation} 

The full-row rank assumption on $\Psi_{\nu,\mu}$ implies first that $|{\mathcal N}_{\nu,\mu}|\geq n$. For other conditions, 
rewrite (\ref{kapd}) by plugging $t_{i-1}=k_{i-1}$ and changing the variables as $t=k_i-l-1$, we derive
\begin{equation}\label{abi1}
	\kappa_{i-1}(k_{i-1}) = \hat{A}^{\delta_{i-1}(\chi)}_\nu \hat{x}(k_{i-1})+\sum_{t=0}^{\delta_{i-1}(\chi)-1} 
    \hat{A}^t_\nu \hat{B}_\nu \, u(k_i-t-1).
\end{equation}
In (\ref{abi1}), note that $u(k)$ is supported on $[k_{i-1}\;\; k_i)$. As $\delta_{i-1}(\chi) \rightarrow \infty$, 
$\hat{A}^{\delta_{i-1}(\chi)}_\nu \hat{x}(k_{i-1}) \rightarrow 0$ rapidly from the stability part of Assumption~\ref{stabmin} 
and $\kappa_{i-1}(k_{i-1})$ may be approximated by the second term in (\ref{abi1}). Furthermore, if $u(k)$ is a unit intensity 
white-noise, $\kappa_{i-1}(k_{i-1})$, $i\in {\mathcal N}_{\nu\mu}$ are uncorrelated random variables with asymptotic variances 
equal to the controllability Gramian of $(\hat{A}_\nu,\hat{B}_\nu)$. Hence, ${\rm rank}(\Psi_{\nu,\mu})=n$ if $\delta_{i-1}(\chi)$ 
is large enough on ${\mathcal N}_{\nu,\mu}$. The following result furnishes a solution to Problem~\ref{probform1}. 

\begin{theorem}\label{lem2}
	Suppose that $n \geq 2$ and Assumptions~\ref{stabmin}--\ref{mindwell} hold. Then, (\ref{Upsilonnumu}) holds for all pairs 
    $(\nu,\mu)$ in $\mathbb{S}^2$ satisfying ${\rm rank}(\Psi_{\nu,\mu})=n$.
\end{theorem}

Setting $y(k) \equiv x(k)$ and solving  (\ref{Upsilonnumu}) for all pairs $(\nu,\mu) \in \mathbb{S}^2$ with the rank constraint 
${\rm rank}(\Psi_{\nu,\mu})=n$, we obtain all transformations compatible with (\ref{Upsilondef}) and interpolating the input-output 
data for a given $\varphi$ \cite{Verdult&Verhaegen:2004}. The special case studied in \cite{Vidal&Chiuso&Soatto:2002} is recovered 
by  replacing the rank condition in Theorem~\ref{lem2} with the weaker condition $|{\mathcal N}_{\nu,\mu}|= 1$. We will stop here to 
elaborate on Problem~\ref{probform1}, but direct our attention to Problem~\ref{probform2} which requires stronger assumptions on 
$\varphi$. This subject will be treated in the sequel. 

\section{PE hybrid inputs}\label{hybridinpsec}

We begin by deriving some properties of $\Upsilon(\nu,\mu)$. From (\ref{Upsilondef}), 
$\Upsilon(\nu,\mu)=T_\mu^{-1} T_\nu=(T_\nu^{-1} T_\mu)^{-1}=\Upsilon^{-1}(\mu,\nu)$. Note that $\Upsilon(\nu,\nu)=T_\nu^{-1}T_\nu=I_n$. 
Let us develop a formula for $\Upsilon(\nu,\mu)$. Given $\mu_t \in \mathbb{S}$ for all $1\leq t \leq M$, we have from the factorization 
formula   
\begin{equation}\label{facmat}
\Upsilon(\mu_1,\mu) \, \Upsilon(\mu_2,\mu_1) \, \cdots\, \Upsilon(\nu,\mu_M) = T_\mu^{-1} T_{\mu_1} \cdot T_{\mu_1}^{-1} T_{\mu_2} 
\cdots T_{\mu_M}^{-1} T_\nu = T_\mu^{-1} T_\nu =\Upsilon(\nu,\mu).
\end{equation}
A special case is $\nu=1$, $\mu=\sigma$, and $\mu_t=\sigma-t$ for $1 \leq t \leq  \sigma-2$ when $\sigma>2$. Then, $\Upsilon(1,\sigma)$ 
can be written as a product of the $\sigma-1$ matrices $\Upsilon(t,t+1)$, $1 \leq t \leq  \sigma-1$. Even fewer matrices are needed to 
factorize $\Upsilon(\nu,\mu)$ by permitting matrix inversions as well for any pair $(\nu,\mu) \in \mathbb{S}^2$, in sharp contrast with 
the number of the matrices $\Upsilon(\nu,\mu)$ as the pair $(\nu,\mu)$ with $\nu \neq \mu$ varies over $\mathbb{S}^2$ which is 
$\sigma(\sigma-1)$. We are now ready to give a PE definition for hybrid inputs.

\begin{definition}\label{defper}
    Suppose that $n \geq 2$ and Assumptions~\ref{stabmin}--\ref{mindwell} hold. A hybrid input $\omega$ is PE if $T_\nu$, 
    $1 \leq \nu \leq \sigma$ can be identified uniquely up to one similarity transformation from $(u,\varphi,y)$.
\end{definition}

 In Section~\ref{secbasistf}, we will show that if $\omega$ is PE according to Definition~\ref{defper}, then one can recover the input-output 
 map $y_{\mathcal P}$ of the LSS (\ref{ssx})--(\ref{ssy}) which is equivalent to recovering the Markov parameters of (\ref{ssx})--(\ref{ssy}) 
 by a result derived in \cite{Petreczky&Bako:2023}. Then, $\omega$ is PE for the input-output map as per \cite{Petreczky&Bako:2023}. The following lemma
 provides a simple criterion to check if $\omega$ is PE. 
 
\begin{lemma}\label{lem3}
	Suppose that $n \geq 2$, Assumptions~\ref{stabmin}--\ref{mindwell} hold, and ${\rm rank}(\Psi_{t,t+1})=n$ for all 
    $1 \leq t \leq \sigma-1$. Then, $\omega$ is PE.
\end{lemma}

{\em Proof.} The hypothesis implies that $\Upsilon(t,t+1)$, $1\leq t \leq \sigma-1$ are uniquely determined from (\ref{Upsilonnumu}). 
Then, from $\Upsilon(t,t+1)=T_{t+1}^{-1} T_t$ we get $T_{t+1}=T_t \Upsilon^{-1}(t,t+1)$. Thus, 
$T_t=T_1\Upsilon^{-1}(1,2) \cdots \Upsilon^{-1}(t-1,t)$, $t>1$ which suffices to show that Definition~\ref{defper} holds. \qedwhite

Recall that ${\rm rank}(\Psi_{\nu,\mu})=n$ if $\delta_*(\chi)$ is large and $|\mathcal{N}_{\nu,\mu}| \geq n$. Thus, 
${\rm rank}(\Psi_{t,t+1})=n$ if $\delta_*(\chi)$ is large and $\mathcal{P}_t$ is active in at least $n$ segments. A lower bound on 
the total length of these segments is then $O(n \delta_*(\chi))$. Now, from Lemma~\ref{defper} a lower bound on $|\omega|$ is 
$O((\sigma-1) n \delta_*(\chi))$. This lower bound
is actually attained by a hybrid input. In fact, consider $n$ copies of the state transitions 
$1 \rightarrow 2 \, \cdots \, \sigma-1 \rightarrow \sigma$ glued without gaps to form $\varphi$. We record this important result 
as follows.

\begin{theorem}\label{thm3}
	Suppose that $n \geq 2$ and Assumptions~\ref{stabmin}--\ref{locrec} hold. Then, there exists a PE hybrid input satisfying
    Assumption~\ref{mindwell} and $N \leq c \sigma n$ for some constant $c>0$ that does not depend on $N,n,\sigma$. 
\end{theorem}

Once we learn $y_{\mathcal P}$, there is no reason to restrict its domain by Assumption~\ref{mindwell}. In the next section, 
we provide a graph theory explanation as to why estimating $\sigma-1$ pairs of $\Upsilon(\nu,\mu)$ is enough to retrieve the 
rest. This $\sigma-1$ pairs, however, can not be selected at random and has to satisfy certain criteria. Having established 
that, it then follows immediately that $\Upsilon(t,t+1)$, $1 \leq t \leq  \sigma-1$ is a $\sigma-1$ candidate pair set as well.

\section{Relations with the graph theory}\label{graphsec}
Recall that there are $\sigma(\sigma-1)$ pairs $(\nu,\mu) \in \mathbb{S}^2$ with distinct indices $\nu \neq \mu$. Considering inversions 
this number drops to $\sigma (\sigma-1)/2$. Let us consider those pairs with ${\rm rank}(\Psi_{\nu,\mu})=n$. From (\ref{Upsilonnumu}), 
we then calculate $\Upsilon(\nu,\mu)$ uniquely, hence by inversion $\Upsilon(\mu,\nu)$ uniquely, meaning that the order of appearance is 
irrelevant as long as $\nu \neq \mu$ and ${\rm rank}(\Psi_{\nu,\mu})=n$. In this section, we shall answer the question ``What is the 
minimum number of the pairs that must be learned to be able to extract the rest of the pairs?'' Obviously, this is a relevant question 
to the basis transformation since every pair satisfies ${\rm rank}(\Psi_{\nu,\mu})=n$ if $N$ is large enough. To answer this question, 
we represent the underlying setup via an undirected graph named $\mathcal{G}$ where the nodes are the discrete states. For the sake of 
simplicity, we next consider the $\sigma=4$ case as an example; the $\sigma>4$ and $\sigma<4$ cases follow similarly.  

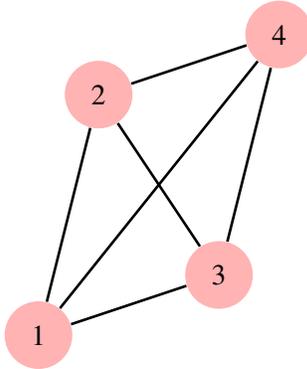
\begin{figure}[hbt!]
\centering
\begin{tikzpicture}
  [scale=.8,auto=left,every node/.style={circle,fill=red!30,minimum size=0.9cm}]
  \node (n1) at (1,1) {1};
  \node (n2) at (2,5)  {2};
  \node (n3) at (4,2)  {3};
\node (n4) at (5,6)  {4};
  \foreach \from/\to in {n1/n2,n2/n3,n3/n4,n2/n4,n1/n3,n1/n4}
      \draw [line width=1pt] (\from) -- (\to);
\end{tikzpicture}
\caption{A graph representation of $\mathcal{G}$.}
	\label{fig100}
\end{figure}

\begin{figure}[hbt!]
    \centering
    \subfloat[\centering \label{fig200}]{{		\begin{tikzpicture}
  [scale=.8,auto=left,every node/.style={circle,fill=red!30,minimum size=0.9cm}]
  \node (n1) at (4,1.2) {1};
  \node (n2) at (1,1)  {2};
  \node (n3) at (5,3.5)  {3};
\node (n4) at (5.2,-2)  {4};
  \foreach \from/\to in {n1/n2,n1/n3,n1/n4}
      \draw [line width=1pt] (\from) -- (\to);
\end{tikzpicture}
	\label{fig200} }}%
    \qquad \qquad
    \subfloat[\centering \label{fig300}]{{\begin{tikzpicture}
  [scale=.8,auto=left,every node/.style={circle,fill=red!30,minimum size=0.9cm}]
  \node (n1) at (3.2,4) {1};
  \node (n2) at (5,1)  {2};
  \node (n3) at (3,-2)  {3};
\node (n4) at (1,1)  {4};
  \foreach \from/\to in {n1/n2,n1/n3,n3/n4}
      \draw [line width=1pt] (\from) -- (\to);
\end{tikzpicture} }}%
    \caption{Two spanning trees for $\mathcal{G}$ given in Figure~\ref{fig100}}%
\end{figure}
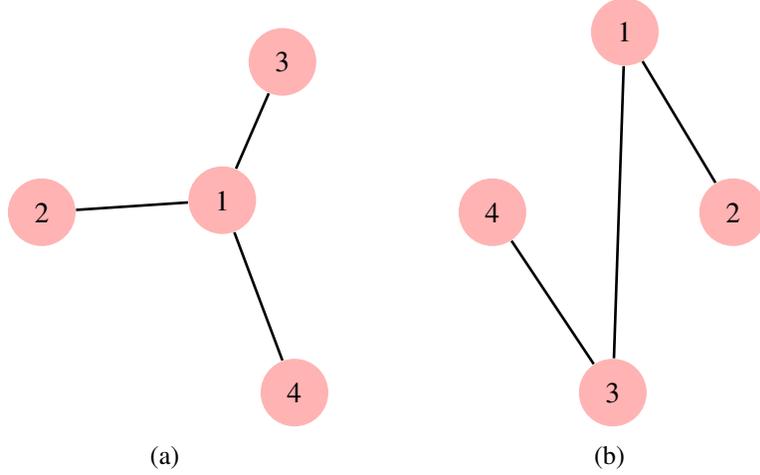

\textbf{Example~1}
Consider the case $\sigma=4$ for which the vertex and the bi-directional edge sets are given by
\begin{equation*}
	\begin{array}{l}
	\mathcal	V = \left\{ {1,2,3,4} \right\},\\
	\mathcal	E = \left\{ {\left( {1,2} \right),\left( {1,3} \right),\left( {1,4} \right),\left( {2,3} \right),
		\left( {2,4} \right),\left( {3,4} \right)} \right\}.
	\end{array}
\end{equation*}
A graph representation of $\mathcal{G}=\left(\mathcal{V}, \mathcal{E} \right)$ is shown in Figure~\ref{fig100}.
Explanation for the graph representation is as follows. Existence of an edge $(\nu,\mu)$ between two nodes $\nu$ and $\mu$ means that 
$\Upsilon(\nu,\mu)$ is determined uniquely. The graph in Figure~\ref{fig100} is complete if there is an edge between any two vertices. 
A graph with this property has $\sigma (\sigma-1)/2$ bi-directional edges. The goal here is to reduce the number of the edges to a 
minimum while keeping the graph connected. We reduce edges in a graph until all edges are of bridge type. A bridge is an edge if removing 
it from a connected graph, renders the graph disconnected. After all reductions, the graph is said a spanning tree which we denote by 
$\mathcal T$. It becomes a spanning tree of $\mathcal G$ when it contains all the vertices originally found in $\mathcal G$. We can 
always find spanning trees for connected graphs \cite{saoub2021graph}. Any spanning tree $\mathcal T$ of $\mathcal G$ has $\sigma$ 
vertices. It is a basic fact that a tree with $\sigma$ vertices has $\sigma-1$ edges. Therefore, to obtain a spanning tree, we follow 
the two basic rules:
\begin{enumerate}[label=(\alph*)]
	\item All vertices from $\mathcal{G}$ are present.
	\item There are exactly $\sigma-1$ edges, all nodes connected, and each node connected to at least one edge.
\end{enumerate}
From the Cayley's formula, the number of the spanning trees for a complete graph with $\sigma$ vertices is $\sigma^{\sigma-2}$. Thus, 
the graph in Figure~\ref{fig100} admits $16$ spanning trees with $3$ bi-directional edges each. We deduce that $\sigma-1$ matrices 
$\Upsilon(\nu,\mu)$ are required with $\nu$ and $\mu$ surjective to $\mathbb{S}$ and there are $\sigma^{\sigma-2}$ clusters of  
$\sigma-1$ matrices. Two examples of spanning trees are shown in Figures~\ref{fig200} --\ref{fig300}. 

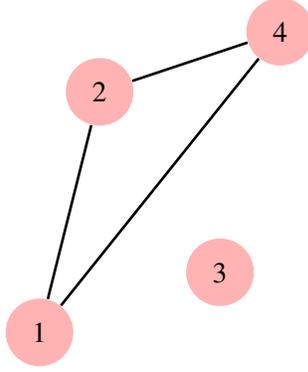
\begin{figure}[hbt!]
	\centering
 \begin{tikzpicture}
  [scale=.8,auto=left,every node/.style={circle,fill=red!30,minimum size=0.9cm}]
  \node (n1) at (1,1) {1};
  \node (n2) at (2,5)  {2};
  \node (n3) at (4,2)  {3};
\node (n4) at (5,6)  {4};
  \foreach \from/\to in {n1/n2,n2/n4,n1/n4}
      \draw [line width=1pt] (\from) -- (\to);
\end{tikzpicture}
	\caption{A graph representation for $\Upsilon^{(2)}$ and the vertex set.}
	\label{fig400}
\end{figure}

We apply the graph theory to the basis construction problem for an LSS with $4$ discrete states. Disregarding order of 
appearances, there are $6$ edges $\Upsilon(\nu,\mu)$. List them all:
\begin{equation*}
	\Upsilon^{(*)}  = \left\{\Upsilon(1,2),\Upsilon(1,3),\Upsilon(1,4),\Upsilon(2,3),\Upsilon(2,4),\Upsilon(3,4)\right\}.
\end{equation*}
The graph representation is shown in Figure~\ref{fig100}. The tree in Figure~\ref{fig200} is the edge subset
\begin{equation*}
\Upsilon^{(1)}=\left\{\Upsilon(1,2),\Upsilon(1,3),\Upsilon(1,4)\right\}
\end{equation*}
with the removed edges retrieved as follows
\begin{equation*}
	\begin{array}{l}
		\Upsilon(2,3) = \Upsilon(1,3)\Upsilon^{-1}(1,2),\\
		\Upsilon(2,4) = \Upsilon(1,4)\Upsilon^{-1}(1,2),\\
		\Upsilon(3,4) = \Upsilon(1,4)\Upsilon^{-1}(1,3).
	\end{array}
\end{equation*}
Similar conclusions are drawn for the spanning tree in Figure~\ref{fig300}. However, if we select 
\begin{equation*}
\Upsilon^{(2)}=\left\{\Upsilon(1,2),\Upsilon(1,4),\Upsilon(2,4)\right\}, 
\end{equation*}
the removed edges are not retrievable although $\Upsilon^{(2)}$ has $\sigma-1=3$ elements. In Figure~\ref{fig400}, $\Upsilon^{(2)}$ 
and the vertex set are plotted. This spanning subgraph is not a spanning tree since it violates the second rule. It is disconnected 
since Node~3 cannot be reached from any other nodes. The $\sigma-1$ edges in Lemma~\ref{lem3} form a spanning tree.  

\section{Basis transformation algorithm}\label{secbasistf}

Assuming that $\omega$ is PE, we define a basis transformation 
$\check{\mathcal P}_\mu =(\check{A}_\mu,\check{B}_\mu,\check{C}_\mu,\check{D}_{\varphi(k_i)})$ where $\mu=\varphi(k_i)$, $\check{A}_\mu=\Upsilon(\mu,1)\hat{A}_\mu \Upsilon^{-1}(\mu,1)$, $\check{B}_\mu=\Upsilon(\mu,1)\hat{B}_\mu$,
$\check{C}_\mu=\hat{C}_\mu \Upsilon^{-1}(\mu,1)$, and $\check{D}_\mu=\hat{D}_\mu$.
We calculate the factors of $\zeta_i(k)$ for all $k \in [k_i,k_{i+1})$ in (\ref{zk}) on setting $\nu=\varphi(k_{i-1})$ and 
starting with
\begin{eqnarray*}
	\hat{C}_\mu \hat{A}^{k-k_i}_\mu T_\mu^{-1} T_\nu &=& \check{C}_\mu \Upsilon(\mu,1)
    \Upsilon^{-1}(\mu,1) \check{A}^{k-k_i}_\mu \Upsilon(\mu,1) T_\mu^{-1} T_\nu \\
	&=& \check{C}_\mu \check{A}^{k-k_i}_\mu T_1^{-1} T_\mu \cdot T_\mu^{-1} T_\nu \\
	&=& \check{C}_\mu \check{A}^{k-k_i}_\mu \Upsilon(v,1).
\end{eqnarray*}
Denote the extended observability matrix of $\check{\mathcal{P}}_\nu$ by $\check{\mathcal{O}}_{i-1}$. From 
$\hat{\mathcal{O} }_{i-1}=\check{\mathcal{O}}_{i-1} \Upsilon (\nu,1)$ and (\ref{hatxk}),
\begin{eqnarray*}\label{hatxk2}
	&{}&\hat{x}(k_{i-1}) = \hat{\mathcal O}^\dag_{i-1}\left(Y_{i-1}-\hat{\Gamma}_{i-1} U_{i-1}\right) \\
	&{}& \hspace{4mm} =\Upsilon^{-1}(\nu,1)\check{O}^\dag_{i-1}\left(Y_{i-1}-\check{\Gamma}_{i-1} U_{i-1}\right)=
	\Upsilon^{-1}(\nu,1) \check{x}(k_{i-1})
\end{eqnarray*}
since $\hat{\Gamma}_{i-1}=\check{\Gamma}_{i-1}$ and hence
\begin{eqnarray*}
 \kappa_{i-1}(k_{i-1}) &=& \hat{A}^{k_i-k_{i-1}}_\nu \hat{x}(k_{i-1})+\sum_{l=k_{i-1}}^{k_i-1} 
 \hat{A}^{k_i-l-1}_\nu \hat{B}_\nu \,  u(l) \\
	&=& \Upsilon^{-1}(\nu,1) \left[ \check{A}_\nu^{k_i-\xi_{i-1}} \check{x}(k_{i-1})+\sum_{l=k_{i-1}}^{k_i-1}
	\check{A}^{k_i-l-1}_\nu \check{B}_\nu \, u(l)\right] \stackrel{\Delta}{=}\Upsilon^{-1}(\nu,1) 
 \check{\kappa}_{i-1}(k_{i-1}).
\end{eqnarray*}
It follows that $\zeta_i(k) = \check{C}_\mu \check{A}^{k-k_i}_\mu \check{\kappa}_{i-1}(k_{i-1})$ for all 
$k \in [k_i,k_{i+1})$. Since $\check{\mathcal P}(k_i) \sim \mathcal{P}(k_i)$ on this interval, Markov parameters are matched 
and from (\ref{zkdef1}) we see that this representation is valid also for $y(k)$ without any dependence on the similarity 
transformations. A pseudo-code implementing the derivations in this section is outlined in Algorithm~1.

\begin{center}   
\begin{tabular}{p{90mm}}\small
	\hrule
	\vspace{0.1cm}
	{\bf Algorithm~1. Basis transformation} 
	\hrule
	\vspace{0.1cm}
	\textbf{Inputs:} $u(k),\varphi(k),y(k)$, $\hat{\mathcal P}_{\varphi(k)} \sim \mathcal{P}_{\varphi(k)}$, $\forall k \in [1\;\;N]$ 					
	\begin{enumerate}
		\item Put $\hat{\mathcal P}_{\varphi(k)}$, $\forall k \in [1\;\;N]$ into $\sigma$-clusters \cite{Esteretal:96} using 
        (\ref{statistic})
	    \item Solve (\ref{Upsilonnumu}), $\forall (\nu,\mu) \in \mathbb{S}^2$ with 
	    ${\rm rank}(\Psi_{\nu,\mu})=n$.
	    \item Use the chain rule to calculate $\Upsilon(j,1)$, $2\leq j \leq \sigma$
	    \item Calculate $\check{\mathcal P}_{\varphi(k_i)}$ for all $i \in [0\;\;i^*]$
	\end{enumerate}			
	\textbf{Outputs:} $(k_i,\check{\mathcal P}(k_i))$ for all $i \in [0\;\;i^*]$.	
	\hrule
\end{tabular}
\end{center}

We state the main result of this paper as follows.

\begin{theorem}
    Suppose $n \geq 2$, Assumptions~\ref{stabmin}--\ref{mindwell} hold, and ${\rm rank}(\Psi_{t,t+1})=n$ for all
    $1 \leq t \leq \sigma-1$. Then, Algorithm~1 solves Problem~\ref{probform2}.
\end{theorem}

The goals set in Section~\ref{basistransform} are achieved by Algorithm~1 as follows. First, if $\hat{\mathcal P}$ were replaced by 
$\tilde{\mathcal P}$, the similarity transformations $\check{T}_\mu$, $1 \leq \nu \leq \sigma$ would of course change. But, these
transformations do not affect the formula $\zeta_i(k) = \check{C}_\mu \check{A}^{k-k_i}_\mu \check{\kappa}_{i-1}(k_{i-1})$.
Therefore, the input-output map is still the same. The reproducibility property follows similarly from the same observation. The third goal
is also achieved since in deriving the formula $\zeta_i(k) = \check{C}_\mu \check{A}^{k-k_i}_\mu \check{\kappa}_{i-1}(k_{i-1})$, no restrictions
were applied to the dwell times.
 
\section{Numerical example}\label{numinsec}

 To illustrate the results derived in this paper, we consider an LSS with three discrete states
 
 \begin{equation*}
 	\begin{array}{l}
 		{A_1} = \left( {\begin{array}{*{20}{c}}{0.32}&{0.31}&0\\{ - 0.32}&{0.31}&0\\0&0&{ - 0.18}\end{array}} \right),
 		{B_1} = \left( {\begin{array}{*{20}{c}}{0.90}&{ - 0.70}\\{0.71}&{ - 0.50}\\{0.80}&{0.47}\end{array}} \right),\\
 		{C_1} =  \left( {\begin{array}{*{20}{c}}{ - 0.55}&{0.20}&{0.80}\\{0.45}&{0.30}&{0.58}\end{array}} \right),\;
 		{D_1} = \left( {\begin{array}{*{20}{c}}{0.90}&{ - 0.70}\\{0.71}&{ - 0.50}\\{0.80}&{0.47}\end{array}} \right);	
 	\end{array}	
 \end{equation*}
 \begin{equation*}
 	\begin{array}{l}
 		{A_2} = \left( {\begin{array}{*{20}{c}}{ - 0.10}&{ - 0.40}&0\\{0.50}&{ - 0.40}&0\\0&0&{0.26}\end{array}} \right),
 		{B_2} = \left( {\begin{array}{*{20}{c}}{0.10}&{ - 0.60}\\{0.32}&{ - 0.66}\\{0.30}&{0.82}\end{array}} \right),\\
 		{C_2} = \left( {\begin{array}{*{20}{c}}{ - 0.80}&{ - 0.10}&{0.70}\\{0.30}&{0.48}&{0.90}\end{array}} \right),
 		{D_2} = \left( {\begin{array}{*{20}{c}}{0.50}&{0.30}\\{ - 0.20}&{ - 0.50}\end{array}} \right);
 	\end{array}
 \end{equation*}
 \begin{equation*}
 	\begin{array}{l}
 		{A_3} = \left( {\begin{array}{*{20}{c}}{0.4}&{0.1}&0\\{0.8}&{0.4}&0\\0&0&{0.8}\end{array}} \right),
 		{B_3} = \left( {\begin{array}{*{20}{c}}{1.5}&{0.9}\\1&{ - 1}\\{ - 1.5}&{2.3}\end{array}} \right),\\
 		{C_3} = \left( {\begin{array}{*{20}{c}}{0.8}&{1.1}&2\\{ - 1.3}&{0.7}&{1.7}\end{array}} \right),
 		{D_3} = \left( {\begin{array}{*{20}{c}}1&0\\1&2\end{array}} \right).
 	\end{array}
 \end{equation*}
 The bimodal LSS formed by the first two discrete states was used in \cite{Pekpe&Mourot&Gasso&Ragot:2004} to numerically investigate 
 performance of a hybrid identification algorithm that merges the Markov parameter based subspace identification and the change detection 
 techniques. The discrete state set for this LSS model comforms with Assumptions~\ref{stabmin}-\ref{mindwell}.
 
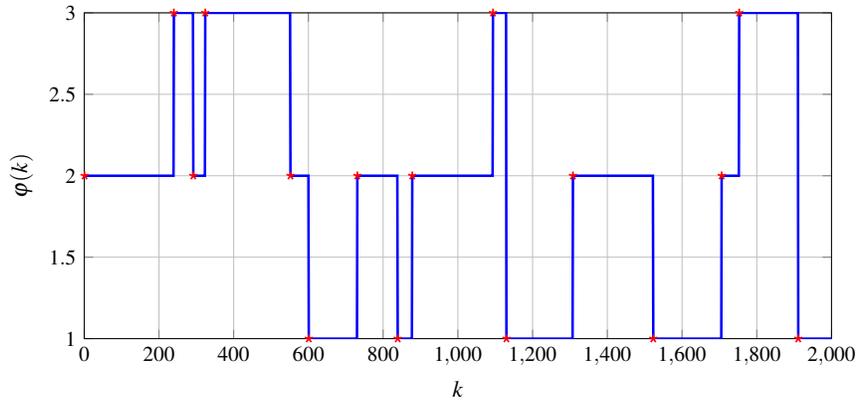
\begin{figure}[hbt!]
    \centering
	\begin{tikzpicture}[scale=0.8]
	\pgfplotsset{every tick label/.append style={font=\small}}
	\begin{axis}
	[xlabel={$k$},
	ylabel={$\varphi(k)$ },
	ylabel near ticks,
	xlabel near ticks,
	xmin=0,
	xmax=2000,
	ymin=1,
	ymax=3,
	yticklabel style={
        /pgf/number format/fixed,
        /pgf/number format/precision=5
},
	grid=major,
    height=7cm,
	width=14cm,
	]
	\foreach \x in {1}{
	\addplot +[blue,mark=none, very thick, ] table[x index=0,y index=\x,col sep=space,]{Plot_data/fig1.txt};
	}	
 	\foreach \x in {1}{
	\addplot +[red,mark=star, thick, only marks, ] table[x index=0,y index=\x,col sep=space]{Plot_data/fig11.txt};
	}	
	\end{axis}
	\end{tikzpicture}	
     \caption{The switching sequence.}
    \label{fig1}
\end{figure}

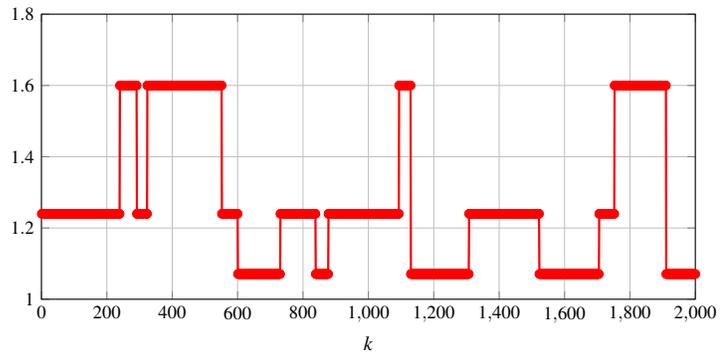
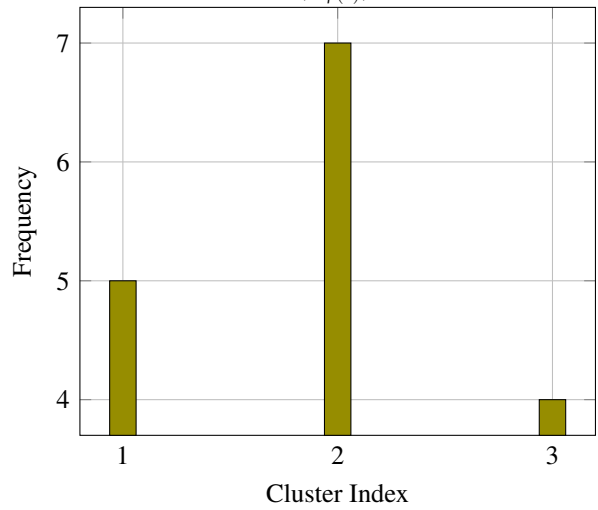
\begin{figure}[hbt!]
        \centering
     \begin{subfigure}[b]{0.5\textwidth}
\begin{tikzpicture}[scale=0.7]
	\pgfplotsset{every tick label/.append style={font=\small}}
	\begin{axis}
	[xlabel={$k$},
	ylabel={},
	ylabel near ticks,
	xlabel near ticks,
	xmin=0,
	xmax=2000,
	ymin=1,
	ymax=1.8,
	yticklabel style={
        /pgf/number format/fixed,
        /pgf/number format/precision=5
},
	grid=major,
    height=7cm,
	width=14cm,
	]
	\foreach \x in {1}{
	\addplot +[red,mark=o, very thick, ] table[x index=0,y index=\x,col sep=space,]{Plot_data/feat.txt};
	}	
	\end{axis}
	\end{tikzpicture}	
 \caption{$\mathcal{M}({\hat A}_{\varphi(k)})$}
 \label{fig11}
     \end{subfigure}
     %\hfill
     \begin{subfigure}[b]{0.5\textwidth}
         \centering
\begin{tikzpicture}
\begin{axis}[
    symbolic x coords={1, 2, 3},
        ylabel = {Frequency},
        xlabel = {Cluster Index},
        grid=major,
    xtick=data,label style={font=\small},tick label style={font=\small} ]
    \addplot[ybar,fill=olive,] coordinates {
        (1,5)
        (2,7)
        (3,4)
    };
\end{axis}
\end{tikzpicture}
     \caption{Histogram of the clusters.}
     \label{fig12}
     \end{subfigure}
        \caption{\small Retrieval of the clusters via the DBSCAN algorithm \cite{Esteretal:96}}.
\end{figure}

 A switching sequence that adheres to the constraint $\delta_*(\chi) \ge n$ (Assumption~\ref{mindwell}) was sampled from a random uniform 
 distribution with its plot given in Figure~\ref{fig1}. 
 We collected the input-output data $(u(k),y(k))$, $k \in [1\;N]$ by feeding the superposed harmonic input signal to the actual LSS. 
 The switching signal and the input signal pair denoted by $\omega$ given in  (\ref{hybrid-input}) were selected to be PE, see 
 Definition~\ref{defper}. We next run the LSS identification algorithm reported in \cite{Bencherki&Turkay&Akcay:2022}. This 
 algorithm delivers $\hat{\mathcal P}_{\varphi(k)}$, $\forall k \in [1\;\;N]$. We run Step~1 of algorithm~1, which will help us 
 retrieve the $\sigma$ clusters. The graph of $\mathcal{M}({\hat A}_{\varphi(k)})$ in Figure~\ref{fig11} reveals three different levels
discerned easily. Figure~\ref{fig12}, likewise, outputs the histogram of clustering. As anticipated, $\sigma$ is correctly
estimated.

\begin{figure}
     \centering
     \begin{subfigure}[b]{0.5\textwidth}
         \centering
\begin{tikzpicture}[scale=0.7]
	\pgfplotsset{every tick label/.append style={font=\small}}
	\begin{axis}
	[xlabel={$k$},
	ylabel={},
	ylabel near ticks,
	xlabel near ticks,
	xmin=0,
	xmax=1000,
	ymin=0,
	ymax=15,
	yticklabel style={
        /pgf/number format/fixed,
        /pgf/number format/precision=5
},
	grid=major,
    height=7cm,
	width=14cm,
 	legend cell align=left,
	legend style={at={(0.82,0.92)},anchor= north west, fill = white}, font = \small,
	legend entries={$h(k,k-1)$\\ $\hat h(k,k-1)$\\},
	legend style={font=\small}
	]
	\foreach \x in {1}{
	\addplot +[blue,mark=o, thick, only marks, ] table[x index=0,y index=\x,col sep=space,]{Plot_data/fig2.txt};
	}	
 	\foreach \x in {1}{
	\addplot +[magenta,mark=star, thick, only marks, ] table[x index=0,y index=\x,col sep=space]{Plot_data/fig22.txt};
	}	
	\end{axis}
	\end{tikzpicture}	
     \end{subfigure}
     \hfill
     \begin{subfigure}[b]{0.5\textwidth}
         \centering
	\begin{tikzpicture}[scale=0.7]
	\pgfplotsset{every tick label/.append style={font=\small}}
	\begin{axis}
	[xlabel={$k$},
	ylabel={},
	ylabel near ticks,
	xlabel near ticks,
	xmin=0,
	xmax=1000,
	ymin=0,
	ymax=9,
	yticklabel style={
        /pgf/number format/fixed,
        /pgf/number format/precision=5
},
	grid=major,
    height=7cm,
	width=14cm,
 	legend cell align=left,
	legend style={at={(0.82,0.92)},anchor= north west, fill = white}, font = \small,
	legend entries={$h(k,k-2)$\\ $\hat h(k,k-2)$\\},
	legend style={font=\small}
	]
	\foreach \x in {1}{
	\addplot +[blue,mark=o, thick, only marks, ] table[x index=0,y index=\x,col sep=space,]{Plot_data/fig3.txt};
	}	
 	\foreach \x in {1}{
	\addplot +[magenta,mark=star, thick, only marks, ] table[x index=0,y index=\x,col sep=space]{Plot_data/fig33.txt};
	}	
	\end{axis}
	\end{tikzpicture}	
     \end{subfigure}
        \caption{\small $h(k,k-1)$ versus $\hat h(k,k-1)$ and $h(k,k-2)$ versus $\hat h(k,k-2)$ before applying the basis correction.}
        \label{fig2}
\end{figure}
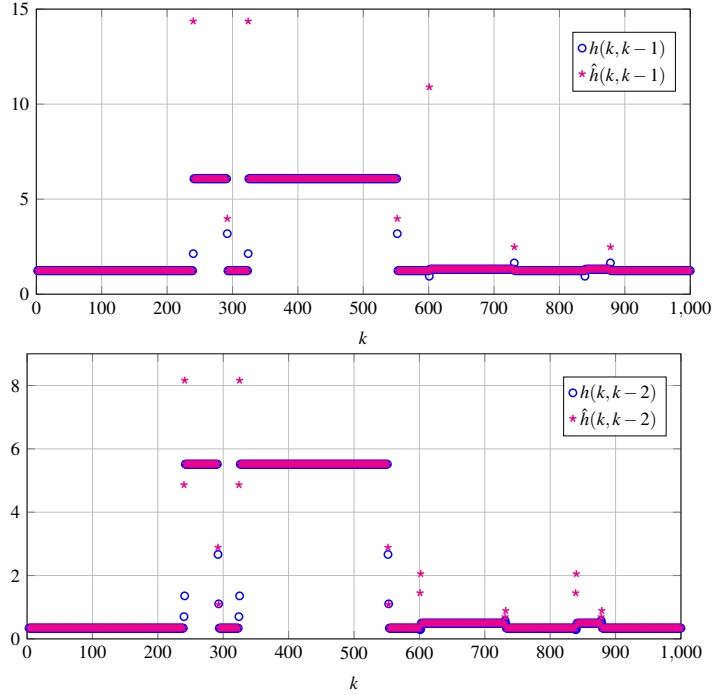

 We pick one index $i$ from each set $\Pi_\nu$, $1\leq \nu \leq \sigma$. Such a representative could be chosen randomly from the class. We end up retrieving the $\sigma$ identified discrete state set. The identified state matrices are 
 \begin{align*}
     \hat A_1=\begin{pmatrix}
   -1.87  &  5.82  & -4.77\\
   -0.9 &   2.07  & -1.25\\
   -0.34  &  0.27  &  0.24
\end{pmatrix},\,
\hat A_2= \begin{pmatrix}
    1.71  & -4.44  &  3.23\\
    0.68 &  -0.59 &  -0.37\\
    0.19  &  0.75  & -1.37
\end{pmatrix},\,
\hat A_3=\begin{pmatrix}
    1.29  & -2.09  &  1.99\\
   -0.46 &   3.42  & -2.88\\
   -0.70 &   3.67  & -3.12
\end{pmatrix}, 
 \end{align*}
 which are the original state transition matrices changed by similarity transformations. The discrete state estimates, if raw used, 
 will not preserve the original LSS input-output map. This could be easily seen by inspecting the original LSS Markov parameters as 
 defined in (\ref{hPhidef}) and the estimated ones defined by 
 \begin{equation*}
	\hat h(k,k)=\hat D_{\varphi(k)}, \qquad \hat h(k,l) =\hat C_{\varphi(k)}\hat\Phi(k,l+1) \hat B_{\varphi(l)}, \;\; l<k \nonumber.
 \end{equation*}
Figure~\ref{fig2} depicts the first two true and estimated Markov parameters, excluding the $h(k,k)$ terms, for $\varphi(k)$ as $k$ 
ranges in the interval $\left[1\;\;1000\right]$. Note the mismatches around the switches. 

\begin{figure}[htb]
    \centering
    \begin{subfigure}[b]{\textwidth}
        \centering
        	\begin{tikzpicture}[scale=0.55]
	\pgfplotsset{every tick label/.append style={font=\small}}
	\begin{axis}
	[xlabel={$k$},
	ylabel={$y_1$ },
	ylabel near ticks,
	xlabel near ticks,
	xmin=0,
	xmax=1000,
	ymin=-30,
	ymax=30,
	yticklabel style={
        /pgf/number format/fixed,
        /pgf/number format/precision=5
},
	grid=major,
    height=7cm,
	width=14cm,
	]
	\foreach \x in {1}{
	\addplot +[blue,mark=none, very thick, ] table[x index=0,y index=\x,col sep=space,]{Plot_data/fig4.txt};
	}	
	\end{axis}
	\end{tikzpicture}	
        %\hfill
        \begin{tikzpicture}[scale=0.55]
	\pgfplotsset{every tick label/.append style={font=\small}}
	\begin{axis}
	[xlabel={$k$},
	ylabel={$e_1$ },
	ylabel near ticks,
	xlabel near ticks,
	xmin=0,
	xmax=1000,
	ymin=0,
	ymax=0.00000000000025,
	yticklabel style={
        /pgf/number format/fixed,
        /pgf/number format/precision=17
},
	grid=major,
    height=7cm,
	width=14cm,
	]
	\foreach \x in {1}{
	\addplot +[red,mark=none, very thick, ] table[x index=0,y index=\x,col sep=space,]{Plot_data/fig44.txt};
	}	
	\end{axis}
	\end{tikzpicture}	
        \caption{The first output $y_1$ on the left and the mismatch error $e_1=\lvert y_1-\hat y_1 \rvert$ on the right.}
    \end{subfigure}
    \vskip\baselineskip
    \begin{subfigure}[b]{\textwidth}
        \centering
        \begin{tikzpicture}[scale=0.55]
	\pgfplotsset{every tick label/.append style={font=\small}}
	\begin{axis}
	[xlabel={$k$},
	ylabel={$y_2$ },
	ylabel near ticks,
	xlabel near ticks,
	xmin=0,
	xmax=1000,
	ymin=-30,
	ymax=30,
	yticklabel style={
        /pgf/number format/fixed,
        /pgf/number format/precision=5
},
	grid=major,
    height=7cm,
	width=14cm,
	]
	\foreach \x in {1}{
	\addplot +[blue,mark=none, very thick, ] table[x index=0,y index=\x,col sep=space,]{Plot_data/fig5.txt};
	}	
	\end{axis}
	\end{tikzpicture}
        \hfill
        \begin{tikzpicture}[scale=0.55]
	\pgfplotsset{every tick label/.append style={font=\small}}
	\begin{axis}
	[xlabel={$k$},
	ylabel={$e_2$ },
	ylabel near ticks,
	xlabel near ticks,
	xmin=0,
	xmax=1000,
	ymin=0,
	ymax=0.0000000000006,
	yticklabel style={
        /pgf/number format/fixed,
        /pgf/number format/precision=17
},
	grid=major,
    height=7cm,
	width=14cm,
	]
	\foreach \x in {1}{
	\addplot +[red,mark=none, very thick, ] table[x index=0,y index=\x,col sep=space,]{Plot_data/fig55.txt};
	}	
	\end{axis}
	\end{tikzpicture}	
        \caption{The first output $y_1$ on the left and the mismatch error $e_2=\lvert y_2-\hat y_2 \rvert$ on the right.}
    \end{subfigure}
    \caption{The true outputs and the mismatch errors.}
    \label{fig6}
\end{figure}
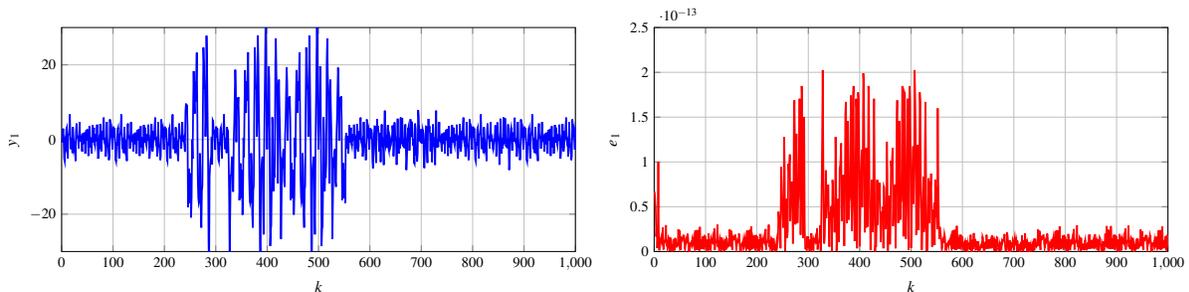
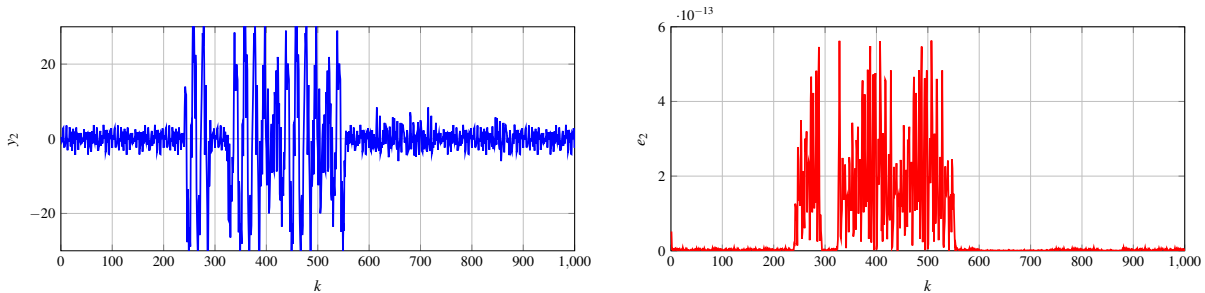

The switching sequence satisfies 
\begin{align*}
    	{\mathcal N}_{1,2} &= \left\{6,\,7,\,10,12\right\}\\
        {\mathcal N}_{2,3} &= \left\{1,\,3,\,9,14\right\}\\    
\end{align*}
which allows us to retrieve both $\Upsilon(1,2)$ and $\Upsilon(2,3)$ following Step~2 in Algorithm~1. Applying Step~3 in Algorithm~1, 
we obtain $\Upsilon(j,1)$, $2\leq j \leq 3$ using the inversion and the chain rules via
\begin{align*}
     \Upsilon(2,1)&=\Upsilon^{-1}(1,2)\\
     \Upsilon(3,1)&=\Upsilon^{-1}(1,2)\Upsilon^{-1}(2,3).
\end{align*}
We found
\begin{align*}
    \Upsilon(2,1)=\begin{pmatrix}
    1.26  &  2.39  & -2.10\\
    0.64  &  1.14 &  -0.44\\
    0.20  &  0.56  &  0.22
\end{pmatrix},\qquad
\Upsilon(3,1)=\begin{pmatrix}
   -2.05  & -2.85  &  6.16 \\
    1.69  &  2.56  & -3.55\\
    2.02  &  2.48   & -4.06
\end{pmatrix}.
\end{align*}
We applied ${\Upsilon(2,1)}$ and ${\Upsilon(3,1)}$ to $ {\hat {\mathcal P}_2}$ and $ {\hat {\mathcal P}_3}$, 
respectively, yielding  $\check{\mathcal P}_2$ and  $\check{\mathcal P}_3$ while $\check{\mathcal P}_1=\hat {\mathcal P}_1$. 
This operation brings the basis of $\hat{\mathcal P}_2$ and $\hat{\mathcal P}_3$ to that of  $\hat {\mathcal P}_1$ 
permitting Markov parameter matching and thus input-output map matching. The success of the presented approach is checked by 
monitoring the mismatch 
error between the predicted outputs and the actual ones to fresh superposed harmonic inputs in the range $\left[1\;\;1000\right]$. 
Figure~\ref{fig6} displays the true output signals and the estimation errors side by side. The match between the two is perfect.

\section{Conclusions}\label{consec}

In this paper, we addressed the basis transformation issue for the LSSs from input-output data, a switching sequence, and locally 
identified discrete states which are similar to the true discrete states. We presented an algorithm to compute a transformation
that leaves the input-output map of the LSS invariant assuming that the hybrid inputs are PE, a notion reformulated in the paper. 
The efficacy of the proposed approach was demonstrated through a numerical example. A link between the basis transformation
problem for the LSSs and the graph theory was also put forward. The results in the paper show that the identifiability and the 
persistence of excitation concepts for the LSSs introduced in the literature are intertwined with the basis transformation problem 
studied in the paper.  In particular, we showed that for the subclass of the LSSs with minimal submodels, there are PE hybrid inputs 
satisfying $N \leq c \sigma n$. We expect our results to impact identifiability and PE input design issues in the LSSs.

and reduced realizations and model-reduction of switched linear systems.


\begin{thebibliography}{99}
	
\bibitem{Lennartsonetal:1996}
    B.~Lennartson, M.~Tittus, B.~Egardt, and S.~Pettersson, ``Hybrid systems in process control,'' {\em IEEE Control Systems
    Magazine}, vol.~16, pp.~45--56, 1996.

\bibitem{Carlonietal:2007}
    R.~Carloni, R.~G. Sanfelice, A.~R. Teel, and C.~Melchior, ``A hybrid control strategy for robust contact detection and 
    force regulation,'' in: Proceedings of the {\em American Control Conference}, New York, NY, pages~1461--1466, July 2007.

\bibitem{Schlegletal:2003}
    T.~Schlegl, M.~Buss, and G.~Schmidt, ``A hybrid systems approach toward modeling and dynamical simulation of dexterous
    manipulation,'' {\em IEEE/ASME Transactions on Mechatronics}, vol.~8, pp.~352--361, 2003.

\bibitem{Glover&Lygeros:2004}
    W.~Glover and J.~Lygeros, ``A stochastic hybrid model for air traffic control simulation,'' in: Proceedings of the  
    {\em International Workshop on Hybrid Systems: Computation and Control}, Philadelphia, PA, pages~372--386, March 2004.

\bibitem{Prandini&Hu:2008}
    M.~Prandini and J.~Hu, ``Application of reachability analysis for stochastic hybrid systems to aircraft 
    conflict prediction,'' in: Proceedings of the {\em 47th IEEE Conference on Decision and Control}, Cancun, Mexico, 
    pages~4036--4041, December 2008.

\bibitem{Daectoetal:2014}
    Grace S.~Daecto, M.~Souza, and J.~C.~Geromel, ``Discrete-time switched linear systems state feedback design with 
    application to networked control,'' {\em IEEE Transactions on Automatic Control}, vol. 60, pp. 877--881, 2014.

\bibitem{DePersis&Tesi:2015}
    C.~De~Persis and P.~Tesi, ``Input-to-state stabilizing control under denial-of-service'', {\em IEEE Transactions on 
    Automatic Control}, vol. 60, pp. 2930--2944, 2015. 

\bibitem{Cetinkayaetal:2018}
    A.~Cetinkaya, H.~Ishii, and T.~Hayakawa, ``Analysis of stochastic switched systems with application to networked 
    control under jamming attacks,'' {\em IEEE Transactions on Automatic Control}, vol.~64, pp.~2013--2028, 2018.

\bibitem{Leeetal:2007}
    J.~Lee, S.~Bohacek, J.~P. Hespanha, and K.~Obraczka, ``Modeling communication networks with hybrid systems,'' 
    {\em IEEE/ACM Transactions on Networking}, vol.~15, pp.~630--643, 2007.
    
\bibitem{Hiskens&Pai:2000}
    I.~A. Hiskens and M.~Pai, ``Hybrid systems view of power system modelling,'' in: Proceedings of the {\em IEEE 
    International Symposium on Circuits and Systems}, Geneva, Switzerland, pages~228--231, May 2000.

\bibitem{Murray-Smith:1998}
    R.~Murray-Smith, ``Modelling human control behaviour with context-dependent Markov-switching multiple models,'' 
    {\em IFAC Proceedings Volumes}, vol.~31, pages~461--466, 1998.

\bibitem{Vidal&Ma:2006}
    R.~Vidal and Y.~Ma, ``A unified algebraic approach to 2-d and 3-d motion segmentation and estimation,'' 
    {\em Journal of Mathematical Imaging and Vision}, vol.~25, pp.~403--421, 2006.

\bibitem{Cinquemanietal:2008}
    E.~Cinquemani, A.~Milias-Argeitis, and J.~Lygeros, ``Identification of genetic regulatory networks: A stochastic 
    hybrid approach,'' {\em IFAC Proceedings Volumes}, vol.~41, pages~301--306, 2008.

\bibitem{Sontag:81}
    E.~D.~Sontag, ``Nonlinear regulation: The piecewise linear approach,'' {\em IEEE Transactions on Automatic 
    Control}, vol. 26, pp. 346--358, 1981.

\bibitem{Ferrarietal:2003}
    G.~Ferrari-Trecate, M.~Muselli, D.~Liberati, and M.~Morari, ``A clustering technique for the identification of
    piecesewise affine systems,'' {\em Automatica}, vol. 39, pp. 205--217, 2003.

\bibitem{Vidal&Soatto&Ma&Sastry:2003}
    R.~Vidal, S.~Soatto, Y.~Ma, and S.~Sastry, ``An algebraic geometric approach to the identification of a class of
    linear systems,'' in: Proceedings of the {42nd IEEE Conference on Decision and Control}, Maui, HI, pp. 167--172, 
    December 2003.

\bibitem{Roll&Bemporad&Ljung:2004} 
    J.~Roll, A.~Bemporad, and L.~Ljung, ``Identification of piecewise affine systems via mixed-integer programming,''
    {\em Automatica}, vol. 40, pp. 37--50, 2004.
 
\bibitem{Bemporad&Garulli&Paoletti&Vicino:2005}
    A.~Bemporad, A.~Garulli, S.~Paoletti, and A.~Vicino, ``A Bounded-error approach to piecewise affine system 
    identification,'' {\em IEEE Transactions on Automatic Control}, vol. 50, pp. 1567--1580, 2005.

\bibitem{Bako:2011}
    L.~Bako, ``Identification of switched linear systems via sparse optimization,'' {\em Automatica}, vol. 47, pp. 668--677.

\bibitem{Ozay&Sznaier&Lagoa&Camps:2012} 
    N.~Ozay, C.~Sznaier, C.~M. Lagoa, and O.~I.~Camps, ``A Sparsification Approach to Set Membership Identification of 
    Switched Affine Systems,'' {\em IEEE Transactions on Automatic Control}, vol.~57, pp. 634--648, 2012.

\bibitem{Hojjatinia&Lagoa&Dabbene:2020}
     S.~Hojjatinia, C.~M.~Lagoa, and F.~Dabbene, ``Identification of switched autoregressive exogenous systems from large 
     noisy datasets,'' {\em International Journal of Robust and Nonlinear Control}, vol.~30, pp.~5777--5801, 2020.  

\bibitem{Verdult&Verhaegen:2002}
    V.~Verdult and M.~Verhaegen, ``Subspace identification of multivariable linear parameter-varying systems,'' 
    {\em Automatica}, vol. 38, pp. 805--814, 2002.

\bibitem{Verdult&Verhaegen:2004}
    V.~Verdult and M.~Verhaegen, ``Subspace identification of piecewise linear systems,'' in: Proceedings of the {\em 43rd 
    IEEE Conference on Decision and Control}, Nassau, Bahamas, pages~3838--3843, December 2004.

\bibitem{Verdult&Verhaegen:2009}
    V.~Verdult and M.~Verhaegen, ``Subspace identification of bilinear and LPV systems for open -and closed-loop data,'' 
    {\em Automatica}, vol. 45, pp. 372--381, 2002.

\bibitem{Verhaegen:93}
    M.~Verhaegen, ``Subspace model identification part 3: Analysis of the ordinary output-error state-space model 
    identification algorithm,'' {\em International Journal of Control}, vol. 56, pp. 555--586, 1993.

\bibitem{Verhaegen:94}
    M.~Verhaegen, ``Identification of the deterministic part of MIMO state space models given in innovations form from 
    input-output data,'' {\em Automatica}, vol. 30, pp. 61--74, 1994.

\bibitem{Verhaegen&Verdult:2007}
    M.~Verhaegen and V.~Verdult, {\em Filtering and System Identification: A Least Squares Approach}, Cambridge University
    Press: NY, 2007.

\bibitem{Vidal&Chiuso&Soatto:2002}
    R.~Vidal, A.~Chiuso, and S.~Soatto, ``Observability and identifiability of jump linear systems,'' in: Proceedings of 
    the {\em 41st IEEE Conference on Decision and Control}, Las Vegas, NV, pages 3614--3619, December 2002.

\bibitem{Elhamifiar:2009} 
    E.~Elhamifiar, M.~Petreczky, and R.~Vidal, ``Rank tests for the observability of discrete-time jump linear systems,''
    in: Proceedings of the {\em American Control Conference}, St. Louis, MO, pages 3025--3032, June 2009.

\bibitem{DeSantis&DiBenedetto:2016}
    E.~De~Santis and M.~D.~Di~Benedetto, ``Observability of hybrid dynamical sysyems,'' {\em Foundations and Trends in in 
    Systems and Control}, vol.~3, pp. 363--540, 2016.

\bibitem{Pekpe&Mourot&Gasso&Ragot:2004}
    K.~M. Pekpe, G.~Mourot, K.~Gasso, and J.~Ragot, ``Identification of switching systems using change detection technique in 
    the subspace framework,'' in: Proceedings of the {\em 43rd IEEE Conference on Decision and  Control}, Nassau, Bahamas, 
    pages~3720--3725, December 2004.    

\bibitem{Borges&Verdult&Verhaegen&Botto:2005}
    J.~Borges, V.~Verdult, M.~Verhaegen, and M.~A.~Botto, ``A switching detection method based on projected subspace 
    classification,'' in: Proceedings of the {\em 44th IEEE Conference on Decision and Control, and the European Control 
    Conference}, Seville, Spain, pages 344--349, December 2005.

\bibitem{Borges&Verdult&Verhaegen:2006}
    J.~Borges, V.~Verdult, and M.~Verhaegen, ``Iterative subspace identification of piecewise linear systems,'' {\em IFAC 
    Proceedings Volumes}, vol. 39, pages 368--373, 2006.

\bibitem{Bako&Mercere&Lecoeuche:2009}
    L.~Bako, G.~Merc\`ere, and S.~Leceouche, ``On-line structured identification with application to switched linear systems, 
    {\em International Journal of Control}, vol. 82, pp. 1496--1515, 2009.    

\bibitem{Vidal:2008}
    R.~Vidal, ``Recursive identification of switched ARX systems,'' {\em Automatica}, vol. 44, pp. 2274--2287, 2008.

\bibitem{Heemels&Schutter&Bemporad:2001}
    W.~P.~M.~H.~Heemels, B.~De~Schutter, and A.~Bemporad, ``Equivalence of hybrid dynamical models,'' {\em Automatica}, vol. 37, 
    pp. 1085--1091, 2001.

\bibitem{Weiland&Juloski&Vet:2006}
    S.~Weiland, A.~Lj.~Juloski, and B.~Vet, ``On the equivalence of switched affine models and switched ARX models,''
    in: Proceedings of the {45th IEEE Conference on Decision \& Control}, San Diego, CA, pages 2614--2618, December 2006.

\bibitem{Paoletti&Roll&Garulli&Vicino:2007}
    S.~Paoletti, J.~Roll, A.~Garulli, A.~Vicino, ``Input-output realization of piecewise affine state space models,'' in: 
    Proceedings of the {\em 46th IEEE Conference on Decision and Control}, New Orleans, pages 3164--3169, 2007. 

\bibitem{Petretczky&Toth&Mercere:2017}
    M.~Petreczky, R.~Toth, and G.~Merc\`ere, ``Realization theory for LPV state space representations with affine dependence,''
    {\em IEEE Transactions on Automatic Control}, vol. 62, pp. 4667--4674. 

\bibitem{Petreczky&Bako&Schuppen:2013}   
    M.~Petreczky, L.~Bako, and H.~Schuppen, ``Realization theory of discrete-time linear switched systems, {\em Automatica}, vol. 49,
    pp. 3337--3344, 2013.

\bibitem{Petreczky&Bako:2023}
    M.~Petreczky and L.~Bako, ``On the notion of persistence of excitation for linear switched systems,'' {\em Nonlinear Analysis: Hybrid 
    Systems}, vol. 48, p.~101308, 2023.

\bibitem{Petreczky&Bako&Lecouche&Motchon:2020}  
    M.~Petreczky, L.~Bako, S.~Lecouche, and K.~M.~D.~Motchon, ``Minimality and identifiability of discrete-time switched 
    autoregressive exegeneous systems,'' {\em International Journal of Robust and Nonlinear Control}, vol. 30, pp. 
    5871--5891, 2020.    

\bibitem{Mu&Chen&Cheng&Bai:2022}   
    B.~Mu, T.~Chen, C.~Cheng, and E.~W.~Bai, ``Persistence of excitation condition for identifying swithched linear systems, 
    {\em Automatica}, vol. 137, p.~110142, 2022.

\bibitem{Bencherki&Turkay&Akcay:2022}
    F.~Bencherki, S.~T\"urkay, and H.~Ak\c{c}ay,  ``Observer-based switched-linear system identification,'' 
    arXiv:2107.14571 [eess.SY], revision submitted to IEEE Transactions on Automatic Control, 2022.

\bibitem{Bencherki&Turkay&Akcay:2023}
     F. Bencherki, S. Türkay, and H. Akçay,  ``Realization of multi-input/multi-output switched linear systems from Markov parameters,''
     {\em Nonlinear Analysis: Hybrid Systems}, vol.~48, p. 101311, 2023. 

\bibitem{Bako:2023}
    L.~Bako, ``On sparsity-inducing methods in system identification and state estimation,'' {\em International Journal of Robust and 
    Nonlinear Control}, vol. 33, pp. 177--208, 2023. 

\bibitem{Esteretal:96}
    M.~Ester, H.~P.~Kriegel, J.~Sander, and X.~Xu, ``A density-based algorithm for discovering clusters in large spatial databases 
    with noise,'' in: Proceedings of the {\em 2nd International Conference on Knowledge Discovery and Data Mining}, Portland, OR, 
    pages 226--231, 1996.    

\bibitem{Kailath:1980}
    T.~Kailath, {\em Linear systems}, Prentice-Hall: Englewood Cliffs, NJ, 1980.

\bibitem{Mercere&Bako:2011}
    G.~Merc\`ere and L.~Bako, ``Parameterization and identification of multivariable state-space systems: A  canonical approach,'' 
    {\em Automatica}, vol.~47, pp.~1547--1555, 2011.

\bibitem{saoub2021graph}
    K.~R.~Saoub, {\em Graph Theory: An Introduction to Proofs, Algorithms, and Applications}, Boca Raton: CRC Press, 2021.

\bibitem{Lopes&Ishihara&Borges:2017}
    R.~V. Lopes, J.~Y. Ishihara, and G.~A. Borges, ``Identification of state-space switched linear systems using clustering and hybrid 
    filtering,'' {\em J. Brazilian Soc. Mech. Sci. Eng.}, vol.~39, pp.~565--573, 2017.
		
\end{thebibliography}
\end{document}